\documentclass[twocolumn]{aastex631}

\pdfoutput=1

\def\chandra{{\itshape Chandra\/}}

\def\wise{{\itshape WISE\/}}
\def\galex{{\itshape GALEX\/}}

\def\xmm{{\itshape XMM-Newton\/}}
\def\nustar{{\itshape NuSTAR\/}}

\def\xray{\hbox{X-ray}}

\def\etal{{et\,al.}}

\def\lgoh{$12 + \log({\rm O}/{\rm H})$}
\def\ltsima{$\; \buildrel < \over \sim \;$}
\def\simlt{\lower.5ex\hbox{\ltsima}}
\def\gtsima{$\; \buildrel > \over \sim \;$}
\def\simgt{\lower.5ex\hbox{\gtsima}}
\def\kms{\ifmmode{~{\rm km~s^{-1}}}\else{~km s$^{-1}$}\fi}
\def\lsim{\lower0.3em\hbox{$\,\buildrel <\over\sim\,$}}
\def\gsim{\lower0.3em\hbox{$\,\buildrel >\over\sim\,$}}
\def\msol{$M_\odot$}
\def\h2{H$_2$}
\def\flux{erg~cm$^{-2}$~s$^{-1}$}
\def\lum{erg~s$^{-1}$}

\def\sfr{$M_{\odot}$~yr$^{-1}$}
\def\lxsfru{erg~s$^{-1}$~($M_{\odot}$~yr$^{-1}$)$^{-1}$}

\def\aap{A\&A}

\def\apjl{ApJL}
\def\apjs{ApJS}

\defcitealias{Sim2021}{Sim21}
\defcitealias{Sen2020}{Sen20}
\defcitealias{Leh2021}{L21}
\defcitealias{Pet2004}{PP04}

\usepackage{underscore}
\usepackage{longtable}

\usepackage{apjfonts}
\shorttitle{X-ray Spectra of Low-Metallicity Galaxies}
\shortauthors{Lehmer et al.}

\begin{document}

%
\title{{\bf {\large Elevated Hot Gas and High-Mass X-ray Binary Emission in Low Metallicity 
Galaxies: Implications for Nebular Ionization and 
Intergalactic Medium Heating in the Early Universe}}}

\correspondingauthor{Bret Lehmer}
\email{lehmer@uark.edu}

\author[0000-0003-2192-3296]{Bret~D.~Lehmer}
\affiliation{Department of Physics, University of Arkansas, 226 Physics Building, 825 West Dickson Street, Fayetteville, AR 72701, USA}
\affiliation{Arkansas Center for Space and Planetary Sciences, University of Arkansas, 332 N. Arkansas Avenue, Fayetteville, AR 72701, USA}

\author[0000-0002-2987-1796]{Rafael~T.~Eufrasio}
\affiliation{Department of Physics, University of Arkansas, 226 Physics Building, 825 West Dickson Street, Fayetteville, AR 72701, USA}

\author[0000-0001-8525-4920]{Antara Basu-Zych}
\affiliation{Center for Space Science and Technology, University of
Maryland Baltimore County, 1000 Hilltop Circle, Baltimore, MD 21250, USA}
\affiliation{NASA Goddard Space Flight Center, Code 662, Greenbelt, MD 20771, USA}

\author[0000-0002-9202-8689]{Kristen Garofali}
\affiliation{NASA Goddard Space Flight Center, Code 662, Greenbelt, MD 20771, USA}

\author{Woodrow Gilbertson}
\affiliation{Department of Physics, University of Arkansas, 226 Physics Building, 825 West Dickson Street, Fayetteville, AR 72701, USA}

\author[0000-0003-3374-1772]{Andrei Mesinger}
\affiliation{Scuola Normale Superiore, 56126 Pisa, PI, Italy}

\author{Mihoko Yukita}
\affiliation{NASA Goddard Space Flight Center, Code 662, Greenbelt, MD 20771, USA}
\affiliation{William H. Miller III Department of Physics and Astronomy, Johns Hopkins University, Baltimore, MD 21218, USA}

%
\begin{abstract}
%

High-energy emission associated with star formation has been proposed as a significant source of interstellar medium (ISM) ionization in low-metallicity starbursts and an important contributor to the heating of the intergalactic medium (IGM) in the high-redshift ($z \simgt 8$) Universe. Using \chandra\ observations of a sample of 30 galaxies at $D \approx$~200--450 Mpc that have high specific star-formation rates of 3--9 Gyr$^{-1}$ and metallicities near $Z \approx 0.3 Z_\odot$, we provide new measurements of the average 0.5--8 keV spectral shape and normalization per unit star-formation rate (SFR). We model the sample-combined \xray\ spectrum as a combination of hot gas and high-mass \xray\ binary (HMXB) populations and constrain their relative contributions. We derive scaling relations of $\log L_{\rm 0.5-8 keV}^{\rm HMXB}$/SFR $= 40.19 \pm 0.06$ and $\log L_{\rm 0.5-2 keV}^{\rm gas}$/SFR $= 39.58^{+0.17}_{-0.28}$; significantly elevated compared to local relations. The HMXB scaling is also somewhat higher than $L_{\rm 0.5-8 keV}^{\rm HMXB}$-SFR-$Z$ relations presented in the literature, potentially due to our galaxies having relatively low HMXB obscuration and young and \xray\ luminous stellar populations. The elevation of the hot gas scaling relation is at the level expected for diminished attenuation due to a reduction of metals; however, we cannot conclude that an $L_{\rm 0.5-2 keV}^{\rm gas}$-SFR-$Z$ relation is driven solely by changes in ISM metal content. Finally, we present SFR-scaled spectral models (both emergent and intrinsic) that span the \xray--to--IR band, providing new benchmarks for studies of the impact of ISM ionization and IGM heating in the early Universe.

%
\end{abstract}
%


%
\section{Introduction}\label{sec:intro}
%

The X-ray power output from normal galaxies (i.e., galaxies that do not harbor
luminous active galactic nuclei [AGN]) is dominated by diffuse hot gas and
\xray\ binaries (XRBs), with additional minor contributions from supernovae and
their remnants \citep[see, e.g.,][for reviews]{Fab1989,Fab2006,Fab2019}.  The
{\it Chandra X-ray Observatory} (hereafter, \chandra) has provided resolved
views of the \xray\ emission on subgalactic scales for 100s of relatively
nearby ($D < 30$~Mpc) galaxies, and \xmm\ has provided unprecedented spectral
constraints and physical insight into the emission from several of these
objects.

For star-forming normal galaxies, it has been shown that the galaxy-wide \xray\
power output, $L_{\rm X}$, scales linearly with star-formation rate
\citep[hereafter, $L_{\rm X}$-SFR relations; see,
e.g.,][]{Ran2003,Per2007,Leh2008,Leh2010,Bas2013b,Min2014,Sym2014,Kou2020}.  More
detailed investigations of galaxy components (both from spatial and spectral
analyses) reveal separate linear $L_{\rm X}$-SFR relations for diffuse hot gas
\citep[e.g.,][]{Str2000,Tyl2004,LiW2013a,Min2012b} and high-mass XRBs (HMXBs)
\citep[e.g.,][]{Gri2003,Min2012a,Leh2019}.  However, the levels of
statistical fluctuations in these relations is larger than that expected from
photometric scatter, suggesting that additional physical variations, beyond SFR, influence
$L_{\rm X}$.

With the advent of computationally intensive XRB population synthesis models
\citep[e.g.,][]{Fra2008,Fra2013a,Fra2013b,Lin2010,Mad2017,Wik2017}, it has
become clear that the $L_{\rm X}$-SFR relation for HMXBs should not be
universal, and should have additional non-negligible dependencies on
star-formation history and metallicity.  Metallicity, in particular, is
expected to have a significant impact on the local $L_{\rm X}$-SFR relation and
its scatter, given the breadth of galaxy metallicities present in local-galaxy
samples.  Low metallicity stellar populations are expected to have relatively weak mass
loss from stellar winds, resulting in less angular momentum loss from binary
systems, less binary widening over stellar evolutionary
timescales, and more numerous and massive compact objects.  These ingredients
are predicted to result in an anticorrelation between $L_{\rm X}$/SFR
and metallicity.

One of the key predictions by the \citet{Fra2013a} XRB population synthesis
model is that the typical $L_{\rm X}$(HMXB)/SFR ratio for galaxies in the
Universe should rise with increasing redshift, due to the corresponding decline
in galaxy-average metallicity.  Using stacking analyses in the \chandra\ Deep
Field-South, \citet{Leh2016} showed that there is indeed observable redshift
evolution in the $L_{\rm X}$/SFR relation across the redshift range $z
\approx$~0--2, such that $L_{\rm X}$/SFR~$\propto (1+z)$, consistent with some
of the most viable models from \citet{Fra2013a}.  Additional investigations
have since reached consistent conclusions out to $z \approx$~6
\citep[e.g.,][]{Vit2016,Air2017,Sax2021}.  Studies of local galaxies have
further corroborated the connection between HMXB population formation and
metallicity, showing unambiguously that $L_{\rm X}$(HMXB)/SFR increases with
decreasing metallicity ($Z$), in a claimed $L_{\rm X}$(HMXB)-SFR-$Z$ plane
\citep[see, e.g.,][]{Bas2013a,Bas2016,Pre2013,Dou2015,Bro2016,Kov2020,Leh2021,Vul2021}.
Recently, \citet{For2019,For2020} showed more directly that $z
\approx$~0.1--2.6 galaxies with spectroscopically constrained metallicities
fall onto the $L_{\rm X}$(HMXB)-SFR-$Z$ plane and can explain the observed
redshift evolution.

The recently established connection between the $L_{\rm X}$(HMXB)-SFR relation
and metallicity has led to a surge in interest in HMXBs as potentially
important sources of ionizing radiation in low-metallicity galaxies, and
several observations have found enhanced \xray\ emission in galaxies selected
by extreme ionization signatures, including  Ly$\alpha$ and He~{\small II} emitters,
green peas, and Lyman-continuum galaxies \citep[see,
e.g.,][]{Bro2017,Ble2019,Svo2019,Bay2020,Dit2020,Gro2021}.  Notably, \xray\
emission (particularly from HMXBs) has been implicated as a potential solution
to the long-standing problem of how nebular He~{\small II} is produced in
low-metallicity star-forming galaxies \citep[see, e.g., ][for discussions of
He~{\small II} production in
galaxies]{Pak1986,Sch1996,Shi2012,Jas2013,Cro2016,Sen2019,Ber2021,Oli2021}; however,
uncertainties in the extrapolations of the {\it intrinsic} spectral energy distribution (SED) into the EUV (near the
He~{\small II} ionization energy of 54~eV) has resulted in the connection
between He~{\small II} and \xray\ emission being a topic of active debate
\citep[e.g.,][]{Sch2019,Sen2020,Sax2020,Keh2021,Ric2021,Sim2021}.

In addition to the \xray/extreme-ionization connection, the metallicity-dependent
$L_{\rm X}$-SFR relation has been actively studied for its role in heating the
intergalactic medium (IGM) in the early Universe.
Extrapolation of the most viable \citet{Fra2013b} $L_{\rm X}$(HMXB)/SFR models
to redshifts beyond those constrained by \citet[][$z \approx$~0--2]{Leh2016}
suggests that HMXBs would have dominated the \xray\ emissivity of the
$z\simgt$8 Universe, surpassing the contributions from AGN.  \xray\ emission
from HMXBs is promising as a source of IGM heating, since the emission both
persists on timescales longer than that of ionizing far-UV emission from young
massive stars and traverses longer path lengths before being absorbed
\citep[e.g.,][]{Mir2011,Mad2017}.  Cosmological simulations that track the spin
temperature evolution of the very early Universe ($z \simgt 10$) have
implicated \xray\ emission as potentially the main source of heating the
neutral IGM, prior to reionization \citep[see,
e.g.,][]{Mes2013,Pac2014,Par2019,Eid2020,Hen2020}.  However, similar to studies
of He~{\small II}, the impact of \xray\ heating depends upon knowledge of the
\xray\ spectrum at low energies; in this case, the {\it emergent} \xray\ spectrum at $E
\simlt 1$~keV is of critical importance \citep[e.g.,][]{Pac2014,Das2017}.

As discussed above, the impact of \xray\ emitting components as sources of
ionization and high-redshift IGM heating depend critically on knowledge of the
intrinsic and emergent SEDs from low-metallicity galaxies.  Thus far,
constraints on low-metallicity galaxy \xray\ spectra are based on either small
numbers of objects \citep[e.g., ][]{Thu2004,Leh2015,Gar2020} or shallow
observations of relatively nearby dwarf galaxies \citep[][]{Pre2013,Dou2015},
which are subject to very large statistical scatter in their HMXB
luminosity scaling with SFR \citep[see, e.g.,][]{Leh2021}.  Nonetheless, from
these observations, \citet{Gar2020} showed that the low-metallicity
star-forming galaxies NGC~3310 and VV114 exhibit enhanced power output per SFR
relative to nearly solar-metallicity galaxies, across the full 0.3--30~keV
spectral range, motivating future studies of larger galaxy populations.

Here we provide a new benchmark characterizing how the average \xray\ spectra of
star-forming galaxies scales with SFR for a sample of galaxies selected to have characteristics in common with high-redshift galaxies (e.g., low-metallicity and evidence of active star formation from young stellar populations).
We present \chandra\ observations for a sample of 30 relatively nearby
star-forming-active galaxies, with \lgoh~$\approx$~8.1--8.2 ($Z
\approx$~0.3~$Z_\odot$), SFR~$\approx$~0.5--15~\sfr, and $\log M_\star/M_\odot \approx$~8--9.3.
Our sample size and high SFR values allow us
to provide statistically meaningful constraints on the population-average
0.5--8~keV SED and its statistical scatter.

In \S2, we describe the selection of our low-metallicity galaxy sample.  In
\S3, we describe our multiwavelength FUV to mid-IR SED fitting and present
derived physical properties for our galaxy sample.  In \S4, we describe our new
\chandra\ observations, and discuss our data reduction and \xray\ spectral modeling
procedure.  In \S5, we show modeling of the average \xray\ spectrum scaling with SFR
for our sample, disentangle and constrain HMXB and hot gas contributions to the
overall spectral shape, and investigate the spectra of individual galaxies. In \S6, we study
the $L_{\rm X}$/SFR relation, its statistical scatter, and its dependence on
metallicity for both HMXBs and hot gas.  We further present new constraints and
extrapolations of the emergent and intrinsic average SED of the population,
spanning the mid-IR to X-rays.  Finally, in \S7, we summarize the key results
of the paper.

Throughout this paper, we make reference to \xray\ fluxes that have been
corrected for Galactic absorption, but not intrinsic absorption.  Similarly,
\xray\ luminosities are always reported as ``observed'' quantities that are not
corrected for intrinsic absorption; however, our presentation of intrinsic \xray\ spectra
contain corrections for intrinsic absorption as described in the text.
For comparisons with past studies, we make use of a \citet{Kro2001} initial
mass function (IMF) when deriving physical properties from multiwavelength
UV--to--IR SED fitting, and we utilize a $\Lambda$CDM cosmology, with values of
$H_0$ = 70~\hbox{km s$^{-1}$ Mpc$^{-1}$}, $\Omega_{\rm M}$ = 0.3, and
$\Omega_{\Lambda}$ = 0.7 adopted \citep[e.g.,][]{Spe2003}.

%
%
\begin{figure*}
\centerline{
\hspace{-0.3in}
\includegraphics[width=20cm]{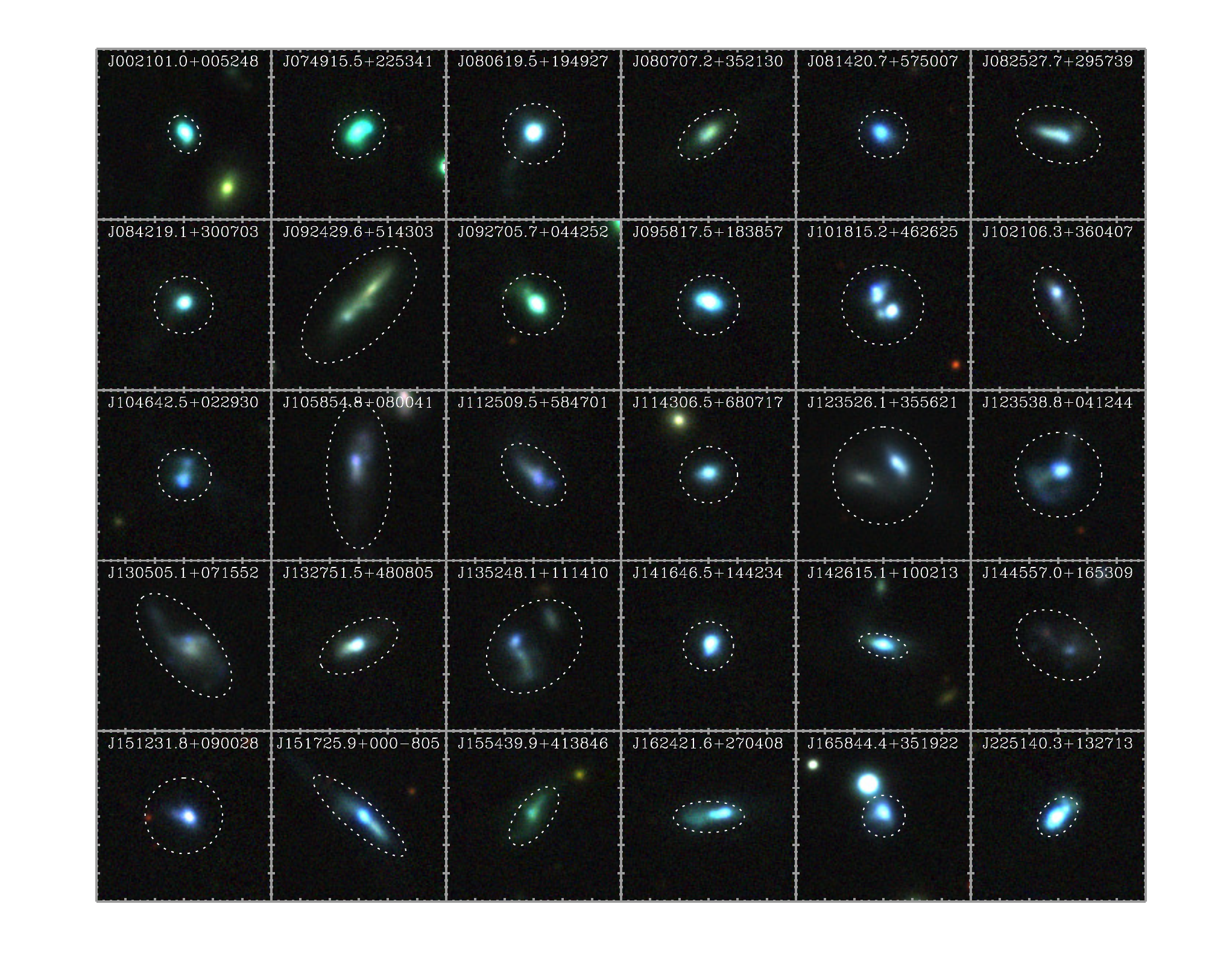}
}
\vspace{-0.45in}
\caption{
PanSTARRS three-band images ($g$ = blue, $i$ = green, and $z$ = red) for the 30
galaxies in our sample.  Each image is 30~$\times$~30~arcsec$^2$ in extent.
The dotted circular or elliptical regions indicate the regions used to extract
\xray\ source counts; the parameters of these regions (i.e., central locations,
major and minor axes, and position angles) are specified in
Table~\ref{tab:props}.
}
\label{fig:img}
\end{figure*}

%
\section{Sample Selection}\label{sec:samp}
%

Our goal was to select a statistically meaningful sample of galaxies that would
allow us to accurately characterize the SFR-scaled \xray\ emission in the
low-metallicity regime.  Such a characterization is critical for developing a
fundamental understanding of the \xray\ heating of the IGM by the first
galaxies and the ionizing emission from \xray\ sources in low-metallicity galaxies.
We started by using the starlight-subtracted emission-line fluxes published in the SDSS DR7 MPA-JHU
value added
catalogs\footnote{\url{http://wwwmpa.mpa-garching.mpg.de/SDSS/DR7/}} to
identify potential AGN and calculate metallicities for a sample of galaxies
that had H$\alpha$, H$\beta$, O~{\small{III}}, and N~{\small II} emission-line
fluxes detected at the $>$3$\sigma$ level.  We removed objects that were
clearly AGN, based on BPT diagnostics; however, we recognize that obscured or
low-luminosity AGN may still remain, and we revisit potential AGN in our sample
based on \xray\ characteristics in \S~\ref{sub:ind}.  For the remaining
galaxies, we used the ``PP04 O3N2'' method \citep[][hereafter, PP04]{Pet2004}
to convert strong emission-line flux ratios into metallicities.  This method
uses an empirical calibration of the OIII]/[H$\beta$]--to--[NII]/[H$\alpha$]
emission-line ratio versus metallicity for $>$100 H~{\small II} regions.  The
conversion has been shown to provide accurate gas-phase metallicities for
galaxies with \lgoh~$> 8.09$ (or $Z \simgt 0.25 Z_\odot$), and it provides one
of the most robust metallicity diagnostics and avoids introducing additional
scatter (up to 0.7 dex) from applying different metallicity methods
\citep[e.g.][]{Kew2008}.  Furthermore, this metallicity calibration allows for
direct comparisons with the \citet[][hereafter, L21]{Leh2021} study of the
metallicity-depdendent HMXB \xray\ luminosity function scaling relation, which
is also based on the PP04 calibration.  \citetalias{Leh2021} provides the most
up-to-date $L_{\rm X}$(HMXB)-SFR-$Z$ relation and quantifies the SFR-dependent
scatter of the relation.

For sample selection purposes, we selected all galaxies with metallicities in
the narrow range of \lgoh~$=$~8.1--8.2 ($Z \approx 0.26$--0.32~$Z_\odot$), near
the lowest limit by which the PP04 diagnostic is accurate.  For context, e.g.,
\citet[][see their Fig.~2]{Mad2017} and \citet{Guo2010} show that a metallicity
of $Z \approx 0.3 Z_\odot$ corresponds to the mass-weighted metallicity of the
Universe at $z \approx 6$ and the metallicity of relatively massive galaxies
$\log M_\star/M_\odot \approx$~9--10 at $z \approx 10$, where the knee of the
stellar mass function is estimated to be $\mathcal{M}_\star^* \approx
10^{9.5}$~\msol\ \citep[e.g.,][]{Ste2021}.  To further identify galaxies that
were more explicitly similar to $z \approx 10$ massive galaxies, we made use of
the SFR and $M_\star$ estimates from \citet{Sal2016}, and filtered our sample
to contain galaxies with SFR~$\approx$~2--20~\sfr\ and $\log M_\star/M_\odot
\approx$~8.5--10 \citep[see, e.g.,][]{Guo2010,Sal2015,Son2016}.  We note that the
properties of these galaxies will not precisely match $z \approx 10$ galaxies
in morphology, stellar density, and star-formation history; however, the three
properties of selection used here (i.e., metallicity, SFR, and $M_\star$) are
expected to be among the most important for XRB formation
\citep[e.g.,][]{Fra2013a,Fra2013b,Wik2017} and can lead to new insight
regarding the \xray\ emission from such galaxy populations.

To optimize our sample for observation, we sorted our initial sample by the
quantity SFR/$d_L^2$, which is a proxy for the \xray\ brightness of a given
galaxy.  We performed simulations to assess how galaxy-to-galaxy scatter
influence the overall average spectral shape and found that constraints based
on a sample of the first 30 galaxies in our sample would give a robust
measurement of the galaxy-population averaged spectrum and SFR scaling with
minimal impact from statistical scatter.  We therefore selected the first 30
galaxies in the sample for observation with \chandra.  The full sample of 30
galaxies was observed by \chandra\ through the combination of archival
observations for three galaxies \citep[J002101.0+005248.1, J080619.5+194927.3,
and J225140.3+132713.4; see][]{Bas2013a} and new observations of the remaining
27 objects (PI: Lehmer).  In Table~\ref{tab:props}, we list the galaxies in our
sample in order of right ascension (R.A.), and in Figure~\ref{fig:img}, we show
corresponding Pan-STARRS $giz$ ({\it blue, green, red\/}) image cutouts.  These
cutouts illustrate that galaxies in our sample have blue optical colors,
clearly indicative of active star formation, and have irregular and complex
morphological types, many of which indicate interacting systems.  These
properties are similar to those of Lyman break analogs (LBAs), which have been
the focus of other \xray\ studies of high-redshift analogs \citep[see,
e.g.,][]{Bas2013a,Bro2016}.

%
\section{Galaxy Physical Properties}\label{sec:phys}
%

Given the morphological variety and multiple components that are apparent for
some of our galaxies, we chose to perform our own detailed photometric extraction and SED fitting of the
available FUV-to-mid-IR data to derive physical properties.  We made use of the
{\ttfamily Lightning} SED fitting code \citep[see][]{Euf2017}, which fits 
broadband SEDs using a non-parametric star-formation history (SFH)
model, including prescriptions for attenuation and nebular
and dust emission \citep[see][for further details]{Doo2021}.  A detailed account of the SED fitting of this sample and its results will be presented in a forthcoming paper (Eufrasio \etal\ in-prep); however, for completeness, we describe the salient details and results below.

We extracted broadband photometry from \galex\ \citep{Mor2007}, SDSS \citep{Ala2015}, PanSTARRS \citep{Cha2016,Wal2020}, 2MASS \citep{twomass}, and
\wise\ \citep{allwise} for all galaxies, providing SED constraints from 0.15--22$\mu$m.  For
all galaxies, we specified either circular or elliptical galactic regions by
visual analysis of the PanSTARRS $g$-band images.  The central positions and
dimensions of these regions are provided in Table~\ref{tab:props} as
Col.(2)--(3) and Col.(6)--(8), respectively, and they are overlaid in
Figure~\ref{fig:img}.  For a given bandbass, we extracted on-source photometry
using expanded versions of these regions, which consisted of the galactic
regions plus four times the half-width at half max PSF appropriate for the
bandpass.  In this process, we subtracted emission from unrelated nearby
Galactic stars (J165844.4+351922) following the procedures in \citet{Euf2017}.
Galactic stars can contribute non-negligible emission in bandpasses at
wavelengths shortward of the \wise\ bands.  Furthermore, in the case of
J105854.8+080041, we found that the relatively large PSF of \wise\ permitted
significant contributions from a nearby unrelated galaxy in the \wise\ bands
that dominated over the target source.   Since these contributions were
inseparable, we chose to model this SED excluding the \wise\ data.

Local backgrounds for all bands were estimated using annuli that were located
2--4 times the expanded regions that were used for on-source photometry.
Within these regions, we adopted the mode of the background distribution of the
pixels as the local background level.  This local background was appropriately
rescaled and subtracted from the on-source photometry.

Using the background-subtracted 0.15--22$\mu$m photometry, we fit each galaxy
SED with {\ttfamily Lightning} using a SFH model that consisted of five
discrete time steps, of constant SFR, at \hbox{0--10~Myr}, 10--100~Myr,
0.1--1~Gyr, 1--5~Gyr, and 5--13.6~Gyr.  The SFH model specifies the intrinsic
stellar and nebular spectral shape, which is further modified by attenuation by
dust and re-emission of the absorbed radiation in the infrared.  We modeled the
attenuation using a three-parameter, modified \citet{Cal2000} extinction curve \citep[as described in][]{Euf2017}, and dust emission was
modeled using \citet{Dra2007} models \citep[with the five-parameter implementation described in][]{Doo2021}.
\color{black}

For each galaxy, we used the derived SFHs to calculate SFR and $M_\star$ values
following:
\begin{equation}\label{eqn:sfr}
{\rm SFR} = \frac{1}{100~{\rm Myr}} \int_0^{\rm 100~Myr} \psi(t) \; dt,
\end{equation}
and
\begin{equation}\label{eqn:mstar}
M_\star = \int_0^{\rm 13.6~Gyr} R(t) \; \psi(t)  \; dt,
\end{equation}
where $t$ represents the look-back time and $\psi(t)$ represents the
instantaneous SFR at look-back time $t$ (e.g., $\psi(0)$ is the instantaneous
SFR at the present day), and $R(t)$ converts the mass of stars formed in the
interval between $t$ and $t + dt$ to its contribution to the present-day
stellar mass at $t=0$.  In Eqn.~\ref{eqn:sfr} (and hereafter), SFR is defined
as the mean value of $\psi(t)$ over the last 100~Myr; this definition is
compatible with those used for widely-used scaling relations that convert
intrinsic UV luminosity to SFR \citep[e.g.,][]{Ken1998,Hao2011}. 

\begin{figure*}
\centerline{
\includegraphics[width=9cm]{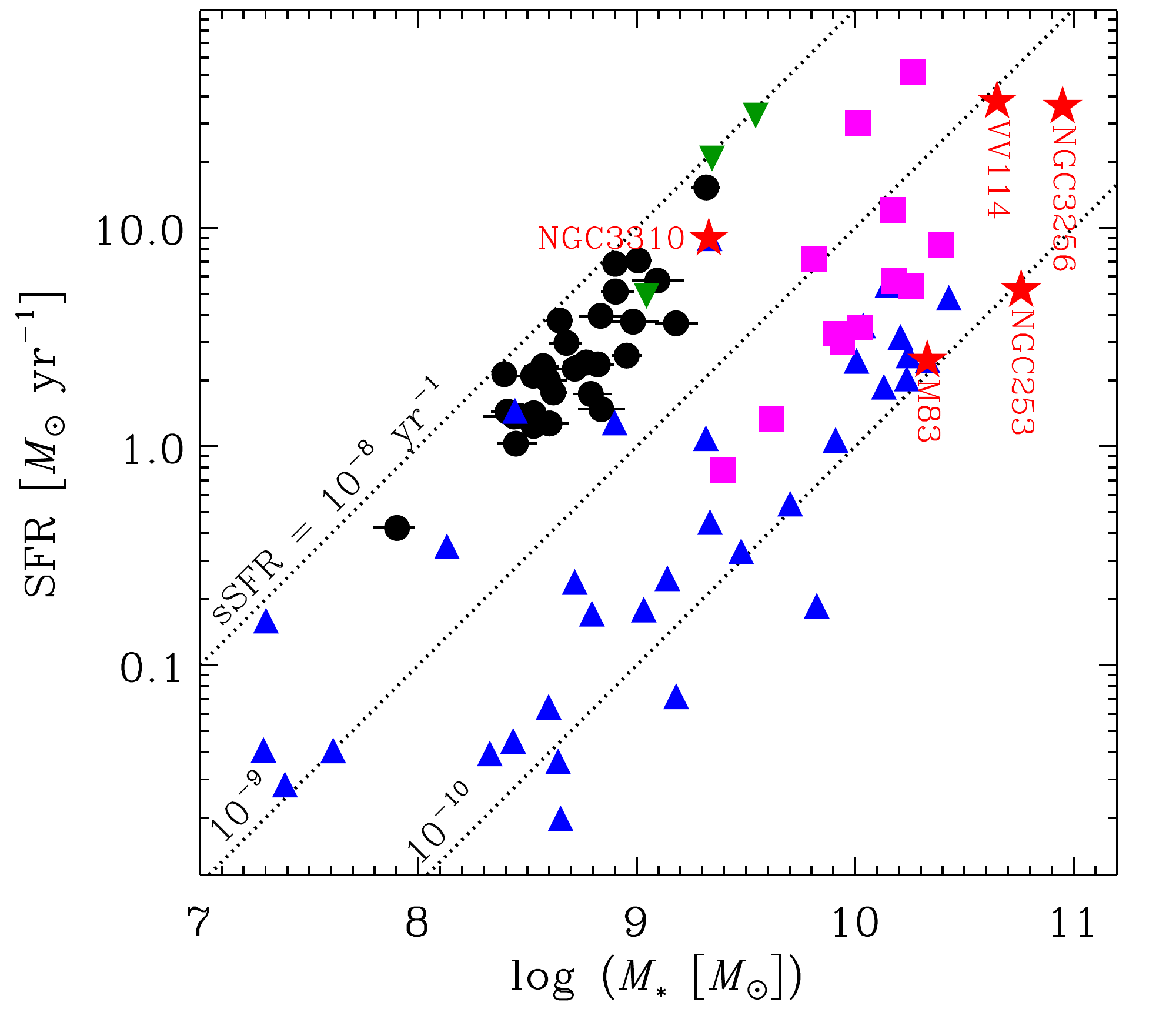}
\hfill
\includegraphics[width=9cm]{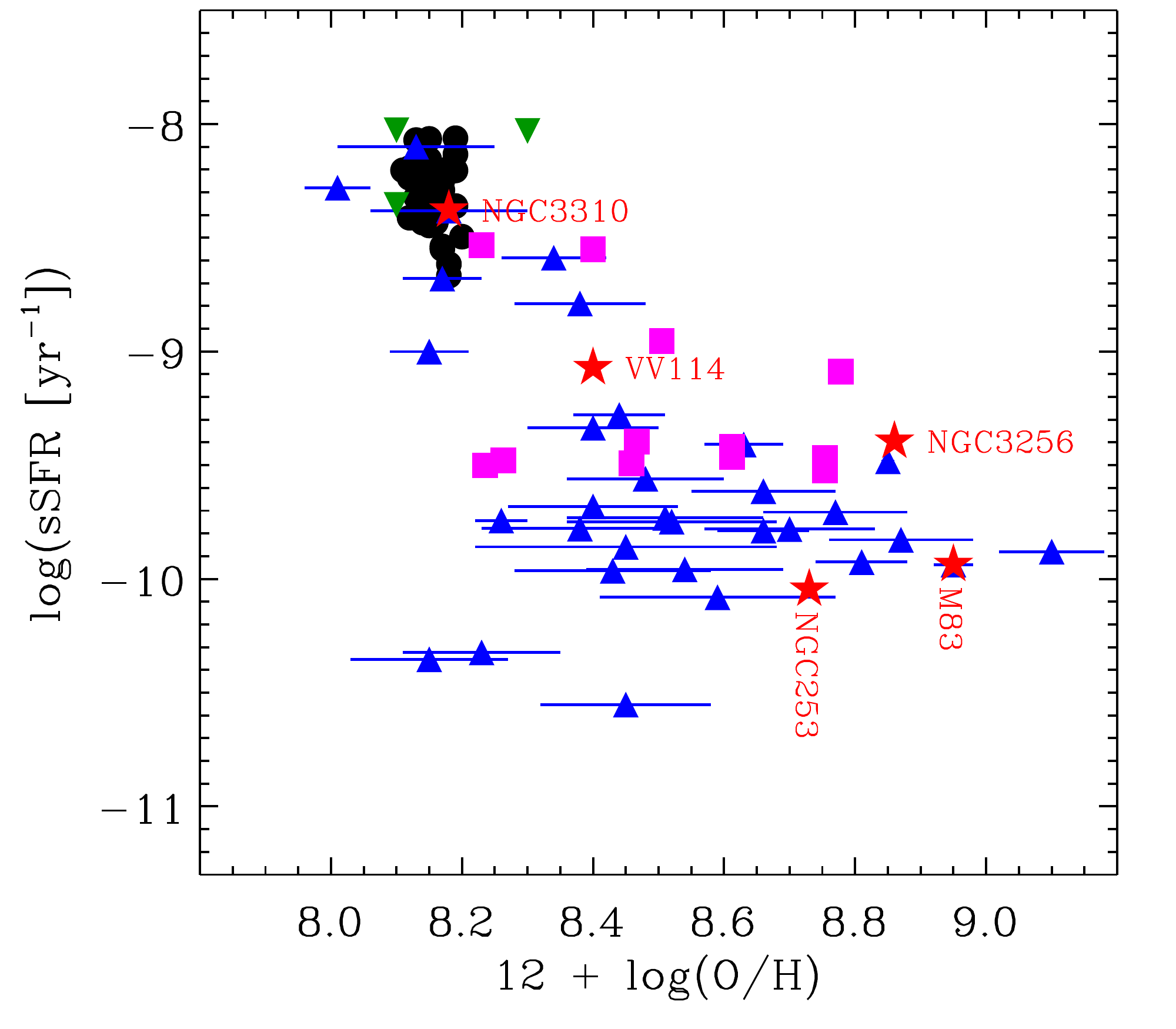}
}
\caption{
($a$) SFR versus logarithm of the stellar mass for the 30 galaxies in our
sample ({\it black circles\/}) and comparison samples from (1)
five well-studied galaxies with spectral constraints across the 0.3--30~keV
band from \citet[][{\it red stars} with annotated labels]{Gar2020}; (2) the
main sample of nearby ($D<30$~Mpc) star-forming galaxies from
\citetalias{Leh2021} ({\it blue triangles\/}); (3) three ``green pea'' galaxies with relatively high SFR that were classified as star-forming galaxies \citep[][{\it green downward triangles\/}]{Svo2019}; (4) the \chandra\ stacked
samples from \citet[][{\it magenta squares}]{For2019,For2020}.
($b$) Logarithm of the specific-SFR (sSFR = SFR/$M_\star$) versus gas-phase
metallicity (\lgoh) for our sample and the comparison samples.  
}
\label{fig:samp}
\end{figure*}

In Figure~\ref{fig:samp}, we show the SFR versus $M_\star$ and specific-SFR (sSFR~$\equiv$~SFR/$M_\star$) versus
metallicity for galaxies in our sample.  By construction, based on the
properties presented in \citet{Sal2016} (see \S~\ref{sec:samp}), the range of
parameters span the area of SFR~$\approx$~\hbox{0.5--15}~\sfr, $\log M_\star/M_\odot
=$~\hbox{8.0--9.3}, and \lgoh~$=$~\hbox{8.1--8.2}.  Figure~\ref{fig:samp} also shows the
properties of relevant comparison samples that have been used to place constraints on \xray\ scaling relations among star-forming galaxies.  These samples include (1) five local galaxies
with excellent global \xray\ spectral constraints from \citet{Leh2015} and
\citet{Gar2020} (NGC~3310, NGC~253, NGC~3256, M83, VV114); (2) the 33
star-forming galaxies presented in the main sample of \citetalias{Leh2021}, who
characterized how the HMXB \xray\ luminosity function (XLF) and $L_{\rm
X}$(HMXB)/SFR scaling varies with metallicity; (3) stacked galaxy subsamples of
$z =$~0.1--2.6 galaxies in the COSMOS and \chandra\ Deep Field-South (CDF-S)
surveys that have been used to directly constrain the $L_{\rm X}$-SFR-$Z$
relation for large galaxy samples \citep{For2019,For2020}; and (4) three green
pea galaxies with high SFRs that have been studied in X-rays by \citet{Svo2019}.

Our sample appears to have SFR, $M_\star$, and metallicity values that overlap NGC~3310, and has very high sSFR values of 3--9~Gyr$^{-1}$, comparable to green peas and extreme emission line galaxies at $z \approx$~7--9 that have intense UV emission from very young stellar populations \citep[e.g.,][]{Mai2017,Mai2018,Sta2017}.  As such, we expect that our galaxies host stellar populations that are young relative to our comparison samples, which span a broader sSFR range.

%
%


\begin{deluxetable*}{lcccccccccc}
\renewcommand\thetable{1}
\tablewidth{1.0\columnwidth}
\tabletypesize{\footnotesize}
\tablecaption{Galaxy Sample Properties}
\tablehead{
\multicolumn{1}{c}{}  & \colhead{} & \colhead{} &\colhead{} & \colhead{} & \multicolumn{3}{c}{\sc Size Parameters} & \colhead{} & \colhead{} & \colhead{} \\
\vspace{-0.25in} \\
\multicolumn{1}{c}{\sc Galaxy} &  \multicolumn{2}{c}{\sc Central Position} & \colhead{$N_{\rm H,gal}$} & \colhead{$D$} & \colhead{$a$} & \colhead{$b$} & \colhead{PA} & \colhead{$\log M_\star$} &  \colhead{SFR} & \colhead{$12 + \log({\rm O/H})$} \\ 
\vspace{-0.25in} \\
\multicolumn{1}{c}{\sc Name} &  \colhead{$\alpha_{\rm J2000}$} & \colhead{$\delta_{\rm J2000}$} & \colhead{(10$^{20}$~cm$^{-2}$)} & \colhead{(Mpc)} & \multicolumn{2}{c}{(arcsec)} & \colhead{(deg)} &  \colhead{($M_\odot$)}   & \colhead{(\sfr)} &  \colhead{(dex)} \\ 
\vspace{-0.25in} \\
\multicolumn{1}{c}{(1)} & \multicolumn{1}{c}{(2)} & \multicolumn{1}{c}{(3)} & \colhead{(4)} & \colhead{(5)} & \colhead{(6)} & \colhead{(7)} & \colhead{(8)} & \colhead{(9)} & \colhead{(10)} & \colhead{(11)} 
}
\startdata
  J002101.0+005248.1 &       00 21 01.0 & +00 52 48.0 & 2.62 &            452.4 &               3.47 & 2.59 &                        26 &    9.32$^{+0.06}_{-0.07}$ &   15.32$^{+0.82}_{-0.67}$ &     8.19 \\
  J074915.5+225342.2 &       07 49 15.5 & +22 53 41.9 & 4.95 &            203.0 &               5.03 & 3.69 &                       128 &    8.62$^{+0.07}_{-0.06}$ &    1.77$^{+0.10}_{-0.10}$ &     8.18 \\
  J080619.5+194927.3 &       08 06 19.5 & +19 49 27.3 & 3.44 &            314.6 &           5.29 &  \ldots  &                    \ldots &    9.01$^{+0.06}_{-0.05}$ &    7.10$^{+0.36}_{-0.36}$ &     8.15 \\
  J080707.2+352130.3 &       08 07 07.2 & +35 21 30.3 & 4.98 &            280.4 &               5.85 & 3.24 &                       127 &    8.84$^{+0.11}_{-0.11}$ &    1.48$^{+0.13}_{-0.17}$ &     8.18 \\
  J081420.8+575008.0 &       08 14 20.7 & +57 50 08.0 & 4.38 &            246.6 &           4.19 &  \ldots  &                    \ldots &    8.53$^{+0.09}_{-0.10}$ &    1.42$^{+0.12}_{-0.14}$ &     8.16 \\
  J082527.7+295739.3 &       08 25 27.7 & +29 57 39.8 & 3.83 &            223.5 &               7.39 & 4.86 &                       254 &    8.60$^{+0.09}_{-0.10}$ &    1.28$^{+0.11}_{-0.11}$ &     8.20 \\
  J084219.1+300703.6 &       08 42 19.1 & +30 07 03.4 & 4.15 &            382.8 &           5.03 &  \ldots  &                    \ldots &    9.18$^{+0.10}_{-0.10}$ &    3.66$^{+0.35}_{-0.48}$ &     8.18 \\
  J092429.9+514301.2 &       09 24 29.6 & +51 43 03.2 & 1.45 &            214.0 &              12.66 & 6.46 &                       137 &    8.53$^{+0.09}_{-0.11}$ &    1.24$^{+0.14}_{-0.16}$ &     8.14 \\
  J092705.7+044251.9 &       09 27 05.7 & +04 42 52.0 & 3.69 &            272.4 &           5.37 &  \ldots  &                    \ldots &    8.95$^{+0.07}_{-0.07}$ &    2.61$^{+0.14}_{-0.18}$ &     8.17 \\
  J095817.5+183858.1 &       09 58 17.5 & +18 38 57.8 & 2.98 &            278.1 &           5.26 &  \ldots  &                    \ldots &    8.90$^{+0.08}_{-0.07}$ &    5.11$^{+0.30}_{-0.29}$ &     8.18 \\
  J101815.1+462623.9 &       10 18 15.2 & +46 26 25.2 & 0.99 &            364.6 &           6.99 &  \ldots  &                    \ldots &    9.09$^{+0.12}_{-0.12}$ &    5.74$^{+0.61}_{-0.42}$ &     8.15 \\
  J102106.4+360408.8 &       10 21 06.3 & +36 04 07.1 & 1.27 &            335.2 &               7.01 & 3.69 &                        22 &    8.77$^{+0.09}_{-0.11}$ &    2.42$^{+0.14}_{-0.19}$ &     8.16 \\
  J104642.5+022930.0 &       10 46 42.5 & +02 29 30.7 & 3.82 &            396.4 &           4.58 &  \ldots  &                    \ldots &    8.68$^{+0.07}_{-0.08}$ &    2.98$^{+0.10}_{-0.11}$ &     8.19 \\
  J105854.8+080044.1 &       10 46 42.5 & +02 29 30.7 & 2.95 &            226.5 &              12.87 & 5.49 &                         0 &    8.44$^{+0.12}_{-0.14}$ &    1.37$^{+0.12}_{-0.12}$ &     8.15 \\
  J112509.5+584700.8 &       11 25 09.5 & +58 47 01.3 & 0.91 &            272.2 &               6.59 & 4.14 &                        45 &    8.41$^{+0.07}_{-0.07}$ &    1.44$^{+0.11}_{-0.11}$ &     8.17 \\
  J114306.5+680717.8 &       11 43 06.5 & +68 07 17.5 & 1.44 &            217.3 &           4.95 &  \ldots  &                    \ldots &    8.53$^{+0.09}_{-0.12}$ &    2.11$^{+0.29}_{-0.22}$ &     8.14 \\
  J123525.9+355622.9 &       12 35 26.1 & +35 56 21.0 & 1.37 &            187.6 &           8.57 &  \ldots  &                    \ldots &    8.59$^{+0.09}_{-0.09}$ &    2.01$^{+0.16}_{-0.15}$ &     8.17 \\
  J123538.7+041244.8 &       12 35 38.8 & +04 12 44.4 & 1.85 &            328.3 &           7.43 &  \ldots  &                    \ldots &    8.83$^{+0.10}_{-0.10}$ &    3.96$^{+0.23}_{-0.25}$ &     8.12 \\
  J130505.1+071552.0 &       13 05 05.1 & +07 15 52.4 & 2.13 &            277.1 &              11.07 & 5.19 &                        40 &    8.72$^{+0.08}_{-0.10}$ &    2.27$^{+0.36}_{-0.21}$ &     8.19 \\
  J132751.6+480805.3 &       13 27 51.5 & +48 08 05.2 & 1.40 &            270.9 &               7.31 & 3.99 &                       119 &    8.79$^{+0.10}_{-0.08}$ &    1.74$^{+0.18}_{-0.16}$ &     8.17 \\
  J135248.3+111410.4 &       13 52 48.1 & +11 14 10.1 & 1.90 &            286.8 &               9.24 & 6.99 &                       135 &    8.46$^{+0.12}_{-0.10}$ &    1.38$^{+0.12}_{-0.12}$ &     8.13 \\
  J141646.5+144235.1 &       14 16 46.5 & +14 42 34.8 & 1.39 &            326.6 &           4.30 &  \ldots  &                    \ldots &    8.65$^{+0.06}_{-0.06}$ &    3.77$^{+0.14}_{-0.10}$ &     8.13 \\
  J142615.1+100213.5 &       14 26 15.1 & +10 02 13.5 & 2.03 &            258.3 &               4.24 & 1.96 &                        75 &    8.39$^{+0.05}_{-0.04}$ &    2.14$^{+0.07}_{-0.06}$ &     8.19 \\
  J144556.9+165308.7 &       14 45 57.0 & +16 53 09.7 & 1.93 &            207.2 &               7.83 & 5.36 &                        57 &    7.90$^{+0.08}_{-0.11}$ &    0.42$^{+0.03}_{-0.03}$ &     8.14 \\
  J151231.8+090028.0 &       14 45 57.0 & +16 53 09.7 & 2.59 &            375.1 &           6.65 &  \ldots  &                    \ldots &    8.90$^{+0.07}_{-0.06}$ &    5.12$^{+0.17}_{-0.21}$ &     8.12 \\
  J151725.9-000805.4 &     15 17 25.9 & $-$00 08 05.1 & 4.74 &            236.4 &              10.31 & 2.74 &                        48 &    8.57$^{+0.07}_{-0.09}$ &    2.33$^{+0.12}_{-0.12}$ &     8.11 \\
  J155440.0+413846.9 &       15 54 39.9 & +41 38 46.7 & 1.60 &            273.7 &               6.02 & 2.92 &                       144 &    8.45$^{+0.09}_{-0.09}$ &    1.03$^{+0.09}_{-0.09}$ &     8.16 \\
  J162421.4+270408.7 &       16 24 21.6 & +27 04 08.1 & 3.53 &            275.3 &               6.08 & 2.73 &                        90 &    8.82$^{+0.07}_{-0.11}$ &    2.37$^{+0.21}_{-0.16}$ &     8.15 \\
  J165844.5+351923.0 &       16 58 44.4 & +35 19 22.7 & 1.82 &            315.4 &           3.65 &  \ldots  &                    \ldots &    8.98$^{+0.12}_{-0.10}$ &    3.72$^{+0.25}_{-0.24}$ &     8.12 \\
  J225140.3+132713.4 &       22 51 40.3 & +13 27 13.7 & 4.85 &            278.7 &               4.09 & 2.76 &                       135 &    8.90$^{+0.04}_{-0.05}$ &    6.85$^{+0.18}_{-0.24}$ &     8.15 \\
\enddata
\tablecomments{Col.(1): Adopted galaxy designation. Col.(2) and (3): Right ascension and declination of the center of the extraction circle or ellipse. Col.(4): Galactic column density based on the {\ttfamily colden} tool in {\ttfamily CIAO}. Col.(5): Adopted distance in units of Mpc.  Col.(6)--(8): Parameters of the circular or elliptical extraction regions, including, respectively, semi-major axis (or radius for the circular case), $a$, semi-minor axis, $b$, and position angle of the semi-major axis east from north, PA. Col.(9) and (10): Logarithm of the galactic stellar mass, $M_\star$, and star-formation rate, SFR, respectively, for the target based on our SED fitting results.  Col.(11): Adopted estimate of the average oxygen abundances, 12+$\log ({\rm O/H})$. For consistency with other studies of XRB scaling relations that include metallicity, we have converted all abundances to the \citet[][PP04]{Pet2004} calibration based on the ratio ([O~{\small III}]$\lambda$5007/H$\beta$)/([N~{\small II}]$\lambda$6584/H$\alpha$). \\
\label{tab:props}}
\end{deluxetable*}

%
\section{X-ray Data Analysis and Spectral Fitting Methodology}
%

\subsection{Data Preparation}\label{sub:data}

Each galaxy was observed with \chandra\ using ACIS-S in a single ObsID.  As
discussed in Section~\ref{sec:samp}, our galaxy sample was selected to contain
the 30 brightest sources (in terms of SFR/$d_L^2$) given our source selection
criteria (i.e., in terms of SFR, $M_\star$, and metallicity). For the 27 newly
observed galaxies, the exposure times were chosen to detect comparable numbers
of source counts for each of the galaxies in our sample and were thus
proportional to $d_L^2$/SFR.  The ObsIDs and exposure times for our sample
galaxies are provided in Col.(2) and (3), respectively, in Table~\ref{tab:fit}.
The exposure times span a range of 11--24~ks, and constitute a total of 555~ks
of \chandra\ observing time.

Our \chandra\ data reduction was carried out using CIAO~v.~4.13 with {\ttfamily
CALDB}~v.~4.9.4.\footnote{http://cxc.harvard.edu/ciao/}  For each ObsID, we
reprocessed pipeline products using the {\ttfamily chandra\_repro} script.  We
removed bad pixels and columns, and filtered the events list and aspect
solutions to include only good time intervals (GTI) without significant ($>$3
$\sigma$) flares above the background level.  The exposure values listed in Col.(3) of
Table~\ref{tab:fit} represent the GTI exposure times that were used in our
analyses.

We next extracted on-source and background spectral (PI) files from each galaxy
using the filtered exposure maps and aspect solutions.  The spectral
extractions were performed using the {\ttfamily specextract} tool using
weighted ancillary response files (ARFs) and redistribution matrix files
(RMFs), since the sources have non-negligible spatial extent.  For our
on-source extractions, we extracted events from the circular or elliptical regions
specified in Table~\ref{tab:props} and displayed in Figure~\ref{fig:img} (see also $\S$\ref{sec:phys} for details).
Background regions were obtained by image inspection in {\ttfamily
ds9},\footnote{https://sites.google.com/cfa.harvard.edu/saoimageds9} in which
we selected 4--6 circular apertures for each galaxy that were located well
outside of (but on the same CCD chip as) the on-source region and were free of
any obvious \xray\ bright sources. The background regions encompassed large
numbers of background counts to ensure high-quality spectral characterization
of the local background.  

In Table~\ref{tab:fit}, we list, in Col.(4), the total extracted 0.5--8~keV
counts from the on-source apertures, $S_{\rm cnts}$, as well as the expected numbers of background
counts for the on-source regions, $B_{\rm cnts}$, in Col.(5).  The background count estimates were
obtained by rescaling the large number of background counts
extracted from the 4--6 background regions (defined above) to the on-source region
areas, after accounting for differences in responses between on-source and background regions.  The estimated net counts (i.e., on-source counts minus estimated background counts) of our sample ranges from $-$3.7 to 51 with a
mean of 8.6 counts per source.  As an ensemble, our sample contains a total of
259 net counts, with 395 on-source counts that contain an estimated 136
background counts.


\begin{deluxetable*}{lccccccccccccccc}
\renewcommand\thetable{2}
\tablewidth{1.0\columnwidth}
\tabletypesize{\footnotesize}
\tablecaption{X-ray Properties of Sample}
\tablehead{
\multicolumn{1}{c}{}  &  \colhead{}  & \colhead{$t_{\rm exp}$} &  \colhead{}  & \colhead{} & \multicolumn{7}{c}{\sc Source Model} & \multicolumn{4}{c}{\sc Global Model} \\
\vspace{-0.25in} \\
\multicolumn{1}{c}{\sc Galaxy} &  \colhead{ObsID} & \colhead{(ks)} & \colhead{$S_{\rm cnts}$} & \colhead{$B_{\rm cnts}$} & \colhead{$A_{\rm cnst}$} & \colhead{$C^{\rm src}$} & \colhead{$C_{\rm exp}^{\rm src}$} & \colhead{$C_{\rm var}^{\rm src}$} & \colhead{$P_{\rm null}^{\rm src}$} & \colhead{$F_{\rm 0.5-8~keV}$} & \colhead{$L_{\rm 0.5-8~keV}$} &  \colhead{$C^{\rm gl}$} & \colhead{$C_{\rm exp}^{\rm gl}$} & \colhead{$C_{\rm var}^{\rm gl}$} & \colhead{$P_{\rm null}^{\rm gl}$} \\
\vspace{-0.25in} \\
\multicolumn{1}{c}{(1)} & \colhead{(2)} & \colhead{(3)} & \colhead{(4)} & \colhead{(5)} & \colhead{(6)} & \colhead{(7)} & \colhead{(8)} & \colhead{(9)} & \colhead{(10)} & \colhead{(11)}  & \colhead{(12)} & \colhead{(13)} & \colhead{(14)} & \colhead{(15)} & \colhead{(16)} 
}
\startdata
  J002101.0+005248.1 & 13014 &    19 &    35 &   1.1 &    1.04$^{+0.20}_{-0.19}$ & 131 & 139 & 334 &                          0.651 &        127.3$^{+24.4}_{-23.1}$ &        311.9$^{+59.8}_{-56.7}$ & 131 & 130 & 333 &           0.952 \\
  J074915.5+225342.2 & 22499 &    20 &     6 &   2.7 &    0.45$^{+0.29}_{-0.22}$ &  42 &  62 & 309 &                          0.244 &         31.8$^{+20.6}_{-15.7}$ &          15.7$^{+10.2}_{-7.7}$ &  45 &  85 & 332 &           0.027 \\
  J080619.5+194927.3 & 13015 &    20 &    35 &   3.6 &    0.95$^{+0.22}_{-0.20}$ & 145 & 144 & 344 &                          0.943 &        110.9$^{+25.6}_{-23.1}$ &        131.4$^{+30.3}_{-27.4}$ & 145 & 141 & 345 &           0.822 \\
  J080707.2+352130.3 & 22512 &    21 &     2 &   3.0 &    0.03$^{+0.34}_{-0.03}$ &  22 &  38 & 276 &                          0.334 &           0.9$^{+10.4}_{-0.9}$ &            0.8$^{+9.8}_{-0.8}$ &  28 &  62 & 309 &           0.055 \\
  J081420.8+575008.0 & 22510 &    23 &     4 &   3.1 &    0.15$^{+0.38}_{-0.15}$ &  40 &  41 & 280 &                          0.995 &           5.6$^{+14.6}_{-5.6}$ &           4.1$^{+10.6}_{-4.1}$ &  45 &  68 & 317 &           0.184 \\
  J082527.7+295739.3 & 22513 &    24 &     7 &   6.1 &    0.21$^{+0.35}_{-0.21}$ &  56 &  64 & 314 &                          0.652 &           8.9$^{+14.6}_{-8.9}$ &            5.3$^{+8.7}_{-5.3}$ &  60 &  91 & 343 &           0.099 \\
  J084219.1+300703.6 & 22505 &    21 &     9 &   4.2 &    1.00$^{+0.55}_{-0.44}$ &  65 &  76 & 326 &                          0.528 &         40.8$^{+22.3}_{-17.8}$ &         71.6$^{+39.1}_{-31.2}$ &  65 &  76 & 326 &           0.543 \\
  J092429.9+514301.2 & 22506 &    21 &    19 &  12.7 &    1.16$^{+0.63}_{-0.52}$ & 112 & 122 & 376 &                          0.604 &         51.5$^{+27.8}_{-23.0}$ &         28.2$^{+15.3}_{-12.6}$ & 113 & 118 & 374 &           0.792 \\
  J092705.7+044251.9 & 22491 &    14 &     8 &   2.7 &    0.89$^{+0.54}_{-0.43}$ &  65 &  63 & 309 &                          0.872 &         51.4$^{+31.2}_{-24.8}$ &         45.6$^{+27.7}_{-22.0}$ &  65 &  66 & 313 &           0.990 \\
  J095817.5+183858.1 & 22487 &    11 &     6 &   2.2 &    0.43$^{+0.33}_{-0.25}$ &  50 &  54 & 298 &                          0.828 &         46.7$^{+36.0}_{-26.9}$ &         43.2$^{+33.3}_{-24.9}$ &  52 &  75 & 322 &           0.197 \\
  J101815.1+462623.9 & 22501 &    21 &    26 &   7.5 &    1.41$^{+0.49}_{-0.42}$ & 162 & 126 & 368 &                          0.061 &         99.3$^{+34.7}_{-29.7}$ &        158.0$^{+55.2}_{-47.2}$ & 163 & 109 & 359 &           0.004 \\
  J102106.4+360408.8 & 22508 &    23 &     6 &   4.8 &    0.32$^{+0.44}_{-0.31}$ &  54 &  56 & 304 &                          0.905 &         11.2$^{+15.6}_{-10.8}$ &         15.1$^{+21.0}_{-14.6}$ &  57 &  77 & 328 &           0.256 \\
  J104642.5+022930.0 & 22494 &    16 &     5 &   2.7 &    1.23$^{+0.77}_{-0.58}$ &  39 &  57 & 303 &                          0.282 &         38.3$^{+24.0}_{-18.1}$ &         72.0$^{+45.2}_{-34.0}$ &  39 &  54 & 300 &           0.368 \\
  J105854.8+080044.1 & 22500 &    20 &    20 &   9.9 &    1.58$^{+0.72}_{-0.61}$ & 127 & 118 & 369 &                          0.664 &         68.9$^{+31.5}_{-26.6}$ &         42.3$^{+19.4}_{-16.4}$ & 128 & 104 & 359 &           0.213 \\
  J112509.5+584700.8 & 22503 &    21 &     5 &   4.3 &    0.58$^{+0.56}_{-0.42}$ &  42 &  59 & 306 &                          0.339 &         18.4$^{+17.9}_{-13.2}$ &         16.4$^{+15.9}_{-11.7}$ &  42 &  69 & 319 &           0.134 \\
  J114306.5+680717.8 & 22509 &    23 &    19 &   4.2 &    1.38$^{+0.42}_{-0.35}$ & 108 & 115 & 355 &                          0.730 &        100.8$^{+30.8}_{-25.9}$ &         57.0$^{+17.4}_{-14.7}$ & 110 &  97 & 345 &           0.474 \\
  J123525.9+355622.9 & 22488 &    12 &    18 &   9.5 &    0.81$^{+0.51}_{-0.41}$ & 119 & 104 & 359 &                          0.446 &         75.4$^{+47.4}_{-38.2}$ &         31.7$^{+19.9}_{-16.1}$ & 119 & 110 & 363 &           0.636 \\
  J123538.7+041244.8 & 22409 &    13 &    18 &   5.2 &    2.32$^{+0.81}_{-0.68}$ & 113 & 109 & 353 &                          0.828 &        139.5$^{+48.6}_{-41.1}$ &        179.9$^{+62.7}_{-53.1}$ & 117 &  78 & 330 &           0.032 \\
  J130505.1+071552.0 & 22502 &    21 &    17 &   9.7 &    1.00$^{+0.57}_{-0.48}$ & 109 & 108 & 362 &                          0.947 &         48.3$^{+27.8}_{-23.2}$ &         44.4$^{+25.6}_{-21.3}$ & 109 & 107 & 362 &           0.921 \\
  J132751.6+480805.3 & 22493 &    16 &     6 &   3.2 &    0.84$^{+0.63}_{-0.48}$ &  45 &  57 & 303 &                          0.509 &         32.6$^{+24.6}_{-18.6}$ &         28.6$^{+21.6}_{-16.3}$ &  45 &  61 & 308 &           0.376 \\
  J135248.3+111410.4 & 22495 &    15 &     5 &   8.7 &    0.39$^{+0.76}_{-0.39}$ &  47 &  69 & 322 &                          0.215 &         10.8$^{+20.8}_{-10.8}$ &         10.6$^{+20.5}_{-10.6}$ &  49 &  84 & 339 &           0.058 \\
  J141646.5+144235.1 & 22498 &    18 &     9 &   2.7 &    0.77$^{+0.43}_{-0.34}$ &  61 &  66 & 313 &                          0.775 &         44.3$^{+24.8}_{-19.4}$ &         56.5$^{+31.7}_{-24.8}$ &  61 &  73 & 321 &           0.511 \\
  J142615.1+100213.5 & 22507 &    23 &    13 &   1.4 &    1.24$^{+0.46}_{-0.38}$ &  88 &  82 & 326 &                          0.739 &         64.9$^{+24.0}_{-19.8}$ &         51.8$^{+19.1}_{-15.8}$ &  88 &  72 & 317 &           0.350 \\
  J144556.9+165308.7 & 22492 &    13 &     9 &   3.7 &    4.52$^{+2.10}_{-1.73}$ &  61 &  75 & 325 &                          0.415 &         72.4$^{+33.6}_{-27.7}$ &         37.2$^{+17.3}_{-14.2}$ &  66 &  50 & 295 &           0.366 \\
  J151231.8+090028.0 & 22497 &    18 &     6 &   5.5 &    0.21$^{+0.35}_{-0.21}$ &  52 &  61 & 309 &                          0.642 &         12.4$^{+20.8}_{-12.4}$ &         20.8$^{+35.0}_{-20.8}$ &  56 &  89 & 340 &           0.076 \\
  J151725.9-000805.4 & 22496 &    17 &    12 &   3.3 &    0.89$^{+0.43}_{-0.36}$ &  89 &  78 & 326 &                          0.554 &         60.5$^{+29.4}_{-24.3}$ &         40.4$^{+19.6}_{-16.2}$ &  88 &  81 & 329 &           0.686 \\
  J155440.0+413846.9 & 22511 &    23 &     5 &   3.0 &    0.38$^{+0.61}_{-0.38}$ &  48 &  42 & 282 &                          0.725 &           8.4$^{+13.6}_{-8.4}$ &           7.6$^{+12.2}_{-7.6}$ &  49 &  57 & 303 &           0.659 \\
  J162421.4+270408.7 & 22489 &    12 &     9 &   1.4 &    1.67$^{+0.75}_{-0.60}$ &  67 &  66 & 311 &                          0.944 &         85.2$^{+38.5}_{-30.8}$ &         77.3$^{+34.9}_{-27.9}$ &  68 &  52 & 296 &           0.350 \\
  J165844.5+351923.0 & 22504 &    22 &     3 &   2.1 &    0.17$^{+0.23}_{-0.15}$ &  26 &  44 & 285 &                          0.280 &          10.3$^{+13.8}_{-9.3}$ &         12.3$^{+16.4}_{-11.1}$ &  33 &  81 & 327 &           0.008 \\
  J225140.3+132713.4 & 13013 &    20 &    53 &   1.7 &    1.35$^{+0.22}_{-0.21}$ & 167 & 173 & 354 &                          0.752 &        194.7$^{+32.4}_{-30.4}$ &        180.9$^{+30.1}_{-28.2}$ & 174 & 144 & 341 &           0.109 \\
\enddata
\tablecomments{Col.(1): Adopted galaxy designation. Col.(2): \chandra\ ObsID. Col.(3): Exposure time in ks.  Col.(4): Total 0.5--8~keV counts extracted from the apertures defined in Table~\ref{tab:props}.  Col.(5): Estimated 0.5--8~keV counts associated with the background (see Section~\ref{sub:data}). Col.(6): Best-fit constant scaling factor, and 16--84\% confidence interval, for the fixed spectral-shape model described in Section~\ref{sub:ind}.  Col.(7): $C$-statistic of the best-fit model.  All models are fit using 512 spectral bins that span the 0.5--8~keV range. Col.(8) and (9): Expected value of the $C$ statistic and its variance, respectively, appropriate for the best-fit model \citep[see methodology in][]{Bon2019}.  Col.(10): Null-hypothesis probability, which we define here as the integral of the $C_{\rm exp}$ distribution from $C$ to $\infty$. Col.(11) and (12): Model 0.5--8~keV fluxes (10$^{-16}$~\flux) and luminosities ($10^{39}$~\lum), along with their 16--84\% confidence intervals.  Col.(13)--(16): $C$ statistic, model-predicted value $C_{\rm exp}$, variance on $C_{\rm exp}$, and null-hypothesis probability for the best-fit global model described in Section~\ref{sub:glo}. \\
}
\label{tab:fit} 
\end{deluxetable*}

\subsection{Spectral Fitting Procedure}\label{sub:fit}

As discussed in \S\ref{sec:intro}, our goals are to both obtain an average
SFR-scaled SED characteristic of the low-metallicity galaxies in our
sample and quantify the scatter in the resulting $L_{\rm X}$/SFR relation.
Since the number of detected counts per source is low, we are unable to
constrain well the \xray\ spectral shapes of individual galaxies.  Therefore, to address
our goals, we start by developing a single ``global'' model that characterizes the
sample-averaged spectral shape and SFR scaling by fitting all data simultaneously.
Next, we fit the \xray\ data for each galaxy individually by fixing the
global-model spectral shape and fitting for a multiplicative renormalization constant
(i.e., a single fitting parameter for each galaxy).

We chose to perform our spectral fitting using {\ttfamily Sherpa} v4.13.0
\citep{Bur2021} with models from {\ttfamily XSPEC} \citep{Arn1996}.  Given the
small numbers of counts for each source, we made use of Poisson statistics in
our spectral analyses using unbinned data.  We limited our data to the
\hbox{0.5--8~keV} energy range to cover where \chandra\ is most sensitive.  Across
this energy range, a given spectral data set (e.g., an on-source extraction
region spectrum) consists of $n_E = 512$ unique spectral channels (or
energies).  
For a given spectral fit, we make use of the Poisson-derived $C$ statistic
\citep{Cas1979,Kaa2017,Bon2019}, which is defined for an individual galaxy as 
\begin{equation}\label{eqn:cstat}
C_j = 2 \sum_{i = 1}^{n_E} M_{ij} - N_{ij} + N_{ij} \ln(N_{ij}/M_{ij}),
\end{equation}
where $M_{ij}$ and $N_{ij}$ are the number of counts for the model and data in
the $i$th energy channel and the $j$th galaxy in the sample.  For our global model, we use the simple summation
\begin{equation}\label{eqn:cglob}
C = \sum_{j=1}^{n_{\rm gal}=30} C_j
\end{equation}
as our model statistic.

For all spectral fits, we started by modeling the local background of each
galaxy independently (see \S\ref{sub:data} for description of background data
extraction).  Our local background model consists of the non-physical
piecewise-linear {\ttfamily CPLINEAR} model at 10 fixed energies that span
0.5--8~keV plus six additional emission lines ({\ttfamily GAUSSIAN})
following \citet{Bar2014}.  These lines have fixed energies at $E
\approx$~\{1.1, 1.5, 1.8, 2.1, 5.9, 7.6\}~keV and line widths spanning
$\sigma_E \approx 10^{-5}$ to 0.2~keV.  Since our goal here is to model the
background shape and normalization, without interest in its physical origins,
we chose to modify the background ARF to be uniform (flat) across all energies.
This choice provides flexibility in the {\ttfamily CPLINEAR} model to match
the observed background spectral shape.  To fit our background model to a given background
data set, we let the normalizations of the 10 energies in the {\ttfamily
CPLINEAR} model and the six emission line intensities vary for the purpose of minimizing
the $C$ statistic (via Eqn~\ref{eqn:cstat}).

For all fits in this paper, we modeled the on-source spectrum as the sum of an
on-source background model plus a galaxy model that is absorbed by a
fixed Galactic column density (Col.(4) in Table~\ref{tab:props}).  For the
on-source background, we fixed all parameters of our background model to their
best-fit values and rescaled the model to the source region using the
{\ttfamily get\_bkg\_scale} method in {\ttfamily Sherpa}.  Thus, the on-source
background component contains no degrees of freedom in our fits.  Given that
\xray\ emission across the 0.5--8~keV range has been observed to contain
significant contributions from hot gas and HMXBs
\citep[e.g.,][]{Min2012b,Pac2014,Leh2015,Smi2018,Smi2019,Gar2020}, we therefore
chose to build our galaxy model as consisting of the sum of hot gas
and HMXB components, with obscuration by Galactic absorption folded
through the on-source response (i.e., the ARF and RMF).

For the hot gas component, previous \xray\ studies of star-forming galaxies
have shown that single or two-temperature thermal plasma models often provide
good fits to diffuse emission observed with \chandra\ and \xmm\ \citep[see,
e.g.,][]{Gri2005,Owe2009,Min2012b,LiW2013a,LiW2013b}.  In a systematic study of the
point-source-excised diffuse emission of 21 nearby star-forming galaxies,
\citet{Min2012b} found that all galaxies in their sample required a $kT=$~0.2--0.3~keV gas component, and 1/3 of the sample required an additional ``hot''
component with $kT=$~0.7--0.8~keV.  

Given the above results from previous studies, we chose to start by utilizing a
two-temperature plasma model component (two {\ttfamily APEC} components) with
Gaussian priors on the temperatures that are based on the results from
\citet{Min2012b}.  Specifically, we implemented priors $kT_1 =$~$0.24\pm
0.05$~keV and $kT_2 =$~$0.71 \pm 0.11$~keV, where the 1$\sigma$ values
represent statistical scatter of the \citet{Min2012b} sample.
Typically the gas components are moderately obscured by the interstellar medium
(ISM) \citep[column density $N_{\rm H} \sim 10^{21}$~cm$^2$; see, e.g.,][]{Min2012b}, and we
model this obscuration using {\ttfamily TBABS}.  We note that the {\ttfamily
TBABS} model assumes solar abundances, and in lower-metallicity environments
like those studied in this paper, the {\ttfamily TBABS} model will
underestimate the true hydrogen column density.  In \S\ref{sub:met}, we discuss
the effects of variable abundances on trends in the emergent hot-gas emission.

Currently, there are few constraints on how the hot gas emission varies with
metallicity.  However, studies of the nearby low-metallicity galaxies NGC~3310
\citep{Leh2015} and VV114 \citep{Gar2020} have shown that the modeled hot gas
temperatures are consistent with those found by \citet{Min2012b}, within
uncertainties, albeit with luminosities per unit SFR ($L_{\rm X}^{\rm
gas}$/SFR) potentially enhanced compared to solar-metallicity galaxies.
Enhanced $L_{\rm X}^{\rm gas}$/SFR is plausibly expected due to relatively low
intrinsic absorption from the low metallicity ISM and/or lower intrinsic column
densities in low-metallicity systems.

For the HMXB model component, we adopted an obscured power-law model
({\ttfamily TBABS$_{\rm HMXB} *$POW$_{\rm HMXB}$}).  
For the HMXB catalog presented in \citetalias{Leh2021}, we find that luminous
sources with $L_{\rm X} \ge 10^{38}$~\lum\ have luminosity-weighted mean column
density $\langle N_{\rm H, HMXB} \rangle = (6 \pm 1) \times 10^{21}$~cm$^2$ and
photon index $\langle \Gamma \rangle = 1.8 \pm 0.1$, where the uncertainties
represent 1$\sigma$ errors on the luminosity-weighted mean values.  We expect
the average spectra of luminous HMXBs to remain consistent with those found in
other galaxies, so we chose to adopt a Gaussian prior with mean and widths
corresponding to the luminosity-weighted mean $\Gamma$ and its 1$\sigma$
uncertainty, respectively.  However, given that our galaxy sample is selected
to be significantly different from typical local galaxies, it is unclear
whether the average absorbing ISM in local galaxies is applicable to this
sample.  We therefore adopted a flat prior on $N_{\rm H, HMXB}$ with a range
from 0 to infinity, and independently compare the recovered average column
density to that found in local galaxies.
In the next section, we outline in detail our global model and present
resulting fits to our data.

%
%
\begin{figure*}
\centerline{
\includegraphics[width=17.5cm]{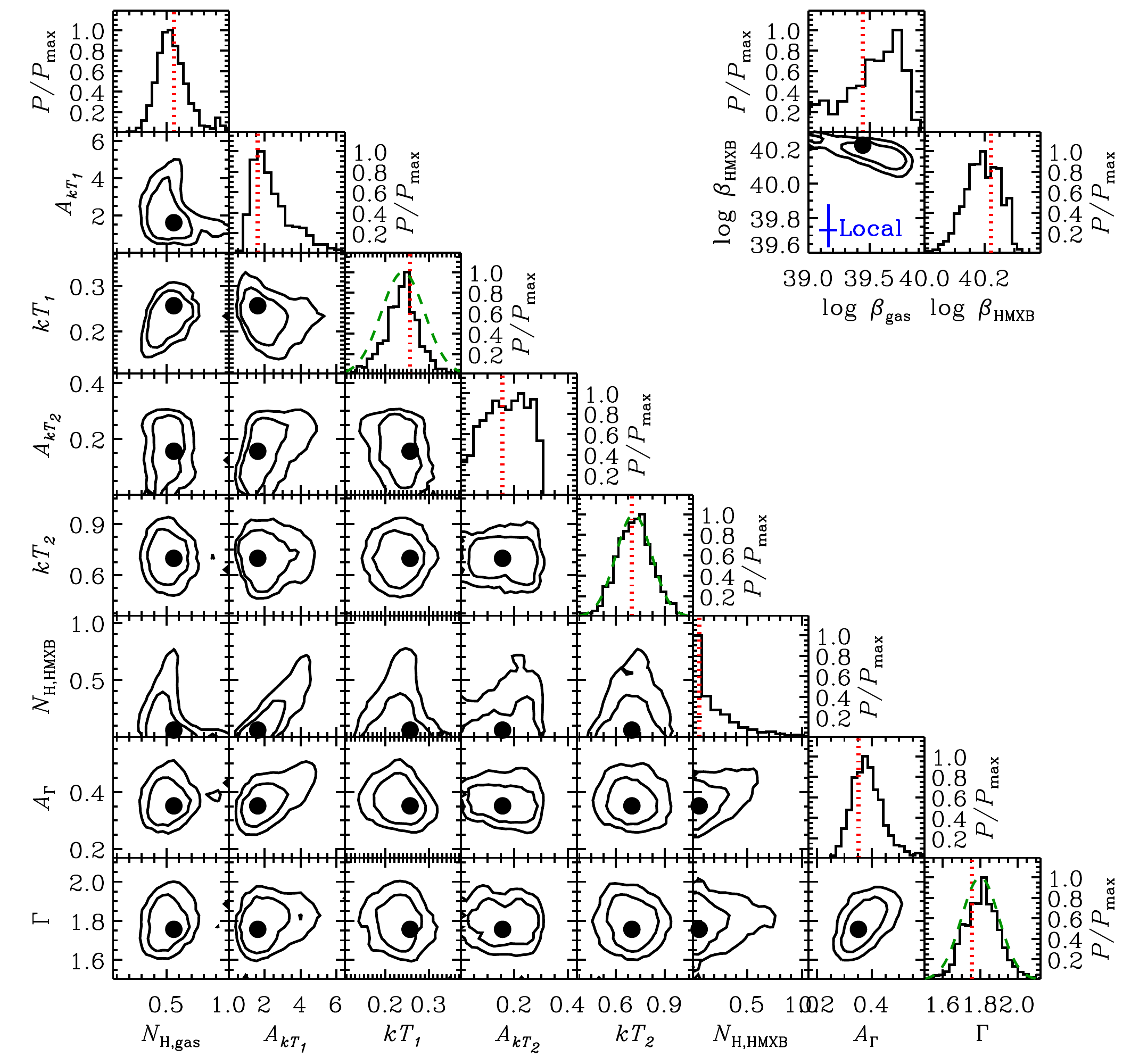}
}
\caption{
Marginalized 1D and 2D probability distribution functions (PDFs) for the eight
parameters in our global model ({\it main corner plot\/}), along with the PDFs
for the model-implied scaling relations $\beta_{\rm HMXB} = L_{\rm
X}$(HMXB)/SFR and $\beta_{\rm gas} = L_{\rm X}$(gas)/SFR ({\it upper right\/}).
Contours represent 68\% and 90\% confidence levels.  The median model parameter
values are shown as vertical dotted lined and filled circles for the 1D and 2D
PDFs.  The adopted prior distributions on $kT_1$, $kT_2$, and $\Gamma$ are
shown as dashed curves in their respective 1D PDF diagrams.  In the panel
showing the 2D PDFs for the scaling relations (i.e., {\it upper right\/}), we
show comparison relations for typical local galaxies from \citet[][hot
gas]{Min2012b} and \citet[][HMXBs]{Leh2019} ({\it blue cross representing
1$\sigma$ uncertainties\/}).  Our model suggests that our sample of
low-metallicity star-forming galaxies have both elevated hot gas and HMXB
scaling relations over those of typical local galaxies.
}
\label{fig:mcmc}
\end{figure*}

%
\section{Results}
%

\subsection{The Global Model}\label{sub:glo}

As discussed in \S\ref{sub:fit}, we first fit a photon-energy ($E$) dependent ``global'' spectral model, $\ell_E$, which we define as the
SFR-normalized intrinsic spectrum in units of
luminosity~per~energy~per~SFR (e.g.,
ergs~s$^{-1}$~keV$^{-1}$~yr~$M_\odot^{-1}$).  As such, we chose to adopt {\tt XSPEC} model
normalizations that are in intrinsic units.  Thus, the $j$th galaxy in our sample will have
a model count-rate spectrum, $\phi_{E,j}$ (uncorrected counts~per~energy~per~second~per~area), and a response-folded
model count spectrum, $S_{E,j}$ (i.e., detected counts per
energy channel), calculated following:
\begin{equation}\label{eqn:phi}
\phi_{E,j} = {\tt TBABS}_{\rm Gal, j}*\left( \frac{{\rm SFR}_j}{4 \pi
d_{L,j}^2} \frac{\ell_E}{E} \right)
\end{equation}
and
\begin{equation}\label{eqn:smod}
S_{E,j} = \left[ {\tt RSP}_j \left( \phi_{E,j} \Delta E \right) + {\tt
URSP}_j\left( {\rm Bkg}_{E,j} \right) \right] \times t_{{\rm exp},j} ,
\end{equation}
where ${\tt TBABS}_{{\rm Gal}, j}$ is the Galactic absorption for the $j$th
galaxy (using the $N_{\rm H, gal}$ values in Col.4 of Table~\ref{tab:props}), $\Delta E$ represents the energy-dependent channel bin width (in energy
units), {\tt RSP}$_j$ and {\tt URSP}$_j$ indicate the use of instrument and
flat (unity) responses for the source and background models, respectively, and $t_{{\rm
exp},j}$ is the exposure time for the $j$th galaxy.  

The intrinsic model consists of the sum of gas and HMXB contributions, $\ell_E
= \ell_E({\rm gas}) + \ell_E({\rm HMXB})$, and can be specified in terms of
{\tt XSPEC} models as:
$$\ell_E({\rm gas}) = E {\tt TBABS}_{\rm gas} * \left({\tt APEC}_{\rm 1} + {\tt APEC}_{\rm 2} \right),$$
\begin{equation}\label{eqn:xrb}
\ell_E({\rm HMXB}) =  E \left({\tt TBABS}_{\rm HMXB} * {\tt POW}_{\rm HMXB}
\right).
\end{equation}
Our global fit contains eight free parameters.  These include (1) the three
normalization terms for the ${\tt APEC}_{\rm 1}$, ${\tt APEC}_{\rm 2}$,
and ${\tt POW}_{\rm HMXB}$ components, $A_{kT_{\rm 1}}$, $A_{kT_{\rm
2}}$, and $A_\Gamma$, respectively, which have flat priors; (2) the ${\tt
TBABS}_{\rm gas}$ and ${\tt
TBABS}_{\rm HMXB}$ absorption components, $N_{\rm H, gas}$ and $N_{\rm H, HMXB}$, respectively, which also have
flat priors; (3) the ${\tt APEC}_{\rm 1}$ and ${\tt APEC}_{\rm 2}$ temperatures, for which we adopt Gaussian priors with mean and standard deviations $kT_1 =$~$0.24\pm
0.05$~keV and $kT_2 =$~$0.71 \pm 0.11$~keV, respectively; and (4) the ${\tt POW}_{\rm HMXB}$ slope, which has Gaussian priors with mean and standard deviation of $\Gamma = 1.8 \pm 0.1$.

To fit the data, we started by minimizing $C$ in Eqn.~\ref{eqn:cglob} using
{\ttfamily Sherpa}'s optimization algorithm with $kT_1$, $kT_2$, and
$\Gamma_{\rm HMXB}$ held fixed at the mean values of their prior distributions.
This provides a set of parameters close to the best-fit solution.
To sample the full posterior distribution function (PDF) of our model
parameters, we utilized a customized adaptive Markov Chain Monte Carlo (MCMC)
procedure.
In this procedure, we incorporated the additional uncertainties on the limited
parameters $kT_1$, $kT_2$, and $\Gamma_{\rm HMXB}$ using their adopted prior
distributions (see above).  Our MCMC algorithm employs the Metropolis Hastings
method \citep{Has1970}, with a vanishing adaptive procedure \citep[see
Algorithm 4 from][]{And2008}.  Model parameters are stepped in accordance with
a covariance matrix.  The covariance matrix is initially set as a diagonal
matrix consisting of the {\ttfamily Sherpa}-derived variances on the five
parameters with flat priors that were intially fit (i.e., $N_{\rm H, gas}$,
$A_{kT_{\rm 1}}$, $A_{kT_{\rm 2}}$, $N_{\rm H, HMXB}$, and $A_\Gamma$), plus
the prior distribution variances for $kT_1$, $kT_2$, and $\Gamma_{\rm HMXB}$.
The covariance matrix is updated at each MCMC step based on the MCMC chain
histories, and then used to direct subsequent MCMC steps.  The algorithm
updates the step sizes in accordance with the covariance matrix until a target
optimal acceptance fraction is achieved \citep{Gel1996}.  To satisfy the
reversibility criterion for Markov Chains, the adaptive aspect of the algorithm
quickly vanishes, and is held fixed for the final 80\% of the MCMC
trials.  The first 20\% of the chains are discarded (i.e., ``burned'') and the
remaining parameter MCMC chains are used to calculate marginalized parameter
distributions.

%
%
\begin{figure*}
\centerline{
\includegraphics[width=9cm]{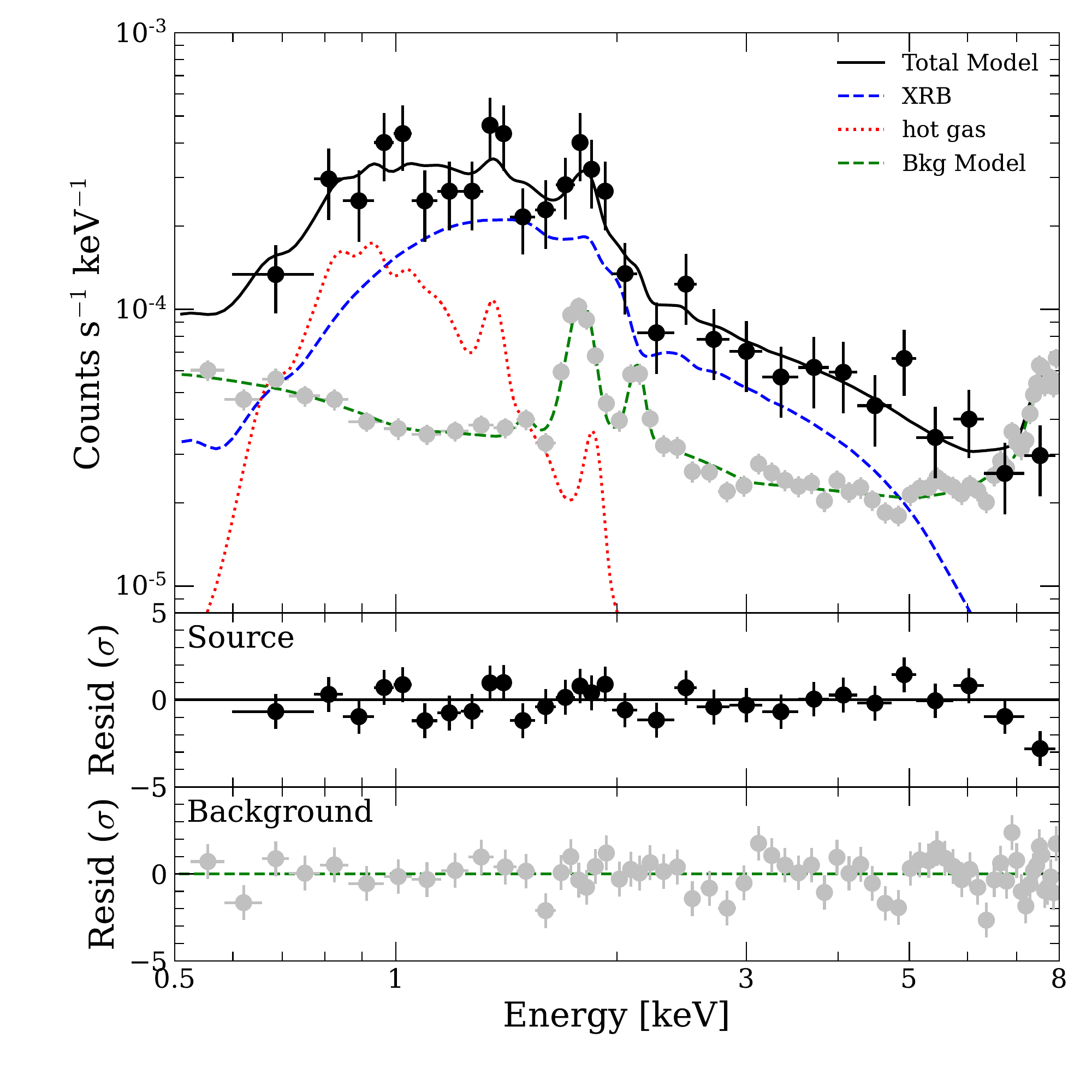}
\hfill
\includegraphics[width=9cm]{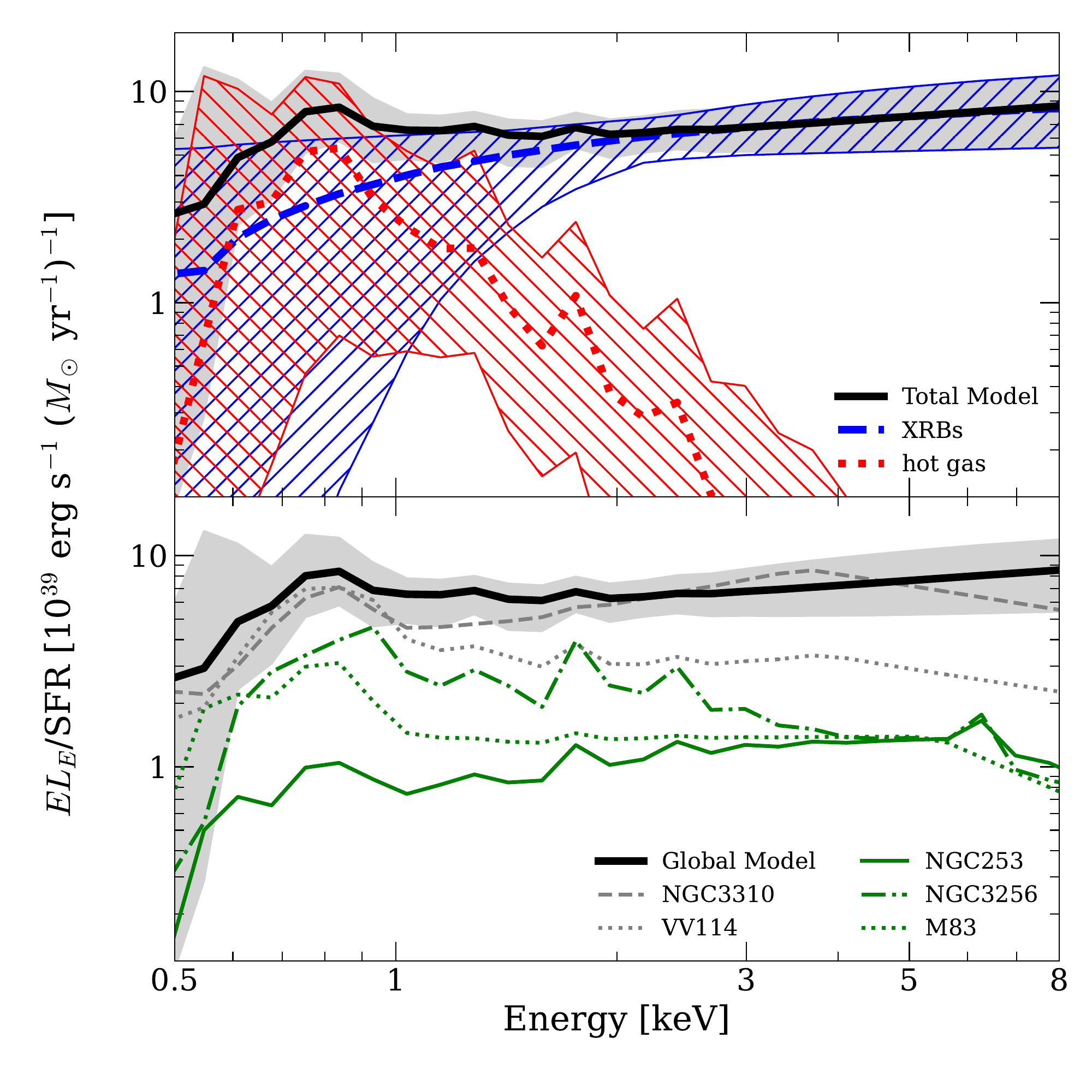}
}
\caption{
({\it Left\/}) Stacked average on-source spectrum for the full sample ({\it
black circles\/}), $\langle \phi_E \rangle$ (see Eqn.~\ref{eqn:phi}), and rescaled background spectrum ({\it gray circles\/}).  For
illustrative purposes, the data have been binned by 12 counts and 120 counts
for the on-source and background spectra, respectively.  Our best-fit global
model is shown as a solid black curve, which contains contributions from HMXBs
({\it blue dashed curve\/}), hot gas ({\it red dotted curve\/}), and background
({\it green dashed curve\/}).  Residuals to the source and background models
are shown in the bottom two panels.
({\it Right\/}) Unfolded global model spectrum, $E L_E$/SFR, and 16--84\% confidence
range ({\it black curve with gray envelope\/}).  In the top panel, we show
model contributions along with their 16--84\% confidence ranges ({\it hatched-shaded regions\/}), and in the bottom panel we compare our SED constraint with
the \nustar\ comparison galaxies.  Our SFR-scaled best-fit spectrum appears elevated
over galaxies with nearly solar metallicity ({\it green curves}; NGC~253,
NGC~3256, and M83) and low-metallicity galaxy VV114 ($\approx$0.3~$Z_\odot$), but similar to the low-metallicity galaxy
NGC~3310 ($\approx$0.2~$Z_{\odot}$).
}
\label{fig:spec}
\end{figure*}

Using an MCMC run of 20,000 trials, we sampled the PDF and identified the model
that most closely maximizes the posterior as our ``best-fit'' global model.  We
note that, given our implementation of non-flat priors, this model is not the
same as the model that minimizes the $C$ statistic.  To test whether our
best-fit global model provides a good fit to the on-source data set for the
full sample, we made use of the methods outlined in \citet{Bon2019} for
calculating the expected value of the $C$ statistic, $C_{\rm exp}$, and its
variance $C_{\rm var}$, which in the limit of large numbers of bins ($\simgt$10
bins) and counts ($\simgt$10 counts) can be taken as a Gaussian distribution
that can be used to test the null hypothesis.  We computed $C_{\rm exp}$ and
$C_{\rm var}$ using Eqn.~11 of \citet{Bon2019} and computed the null hypothesis
probability as
\begin{equation}\label{eqn:pnull}
p_{\rm null} = 1 - {\rm erf}\left( \sqrt{\frac{(C - C_{\rm exp})^2}{2 \; C_{\rm
var}}} \right).
\end{equation}
Under the above definition, a value of $p_{\rm null} \approx 1$ indicates $C
\approx C_{\rm exp}$, while deviations of $C$ away from $C_{\rm exp}$, both low
and high, act to reduce the value of $p_{\rm null}$.  For our global model, we
calculate $p_{\rm null} =$~0.133 (with $C$ being lower than $C_{\rm exp}$),
which indicates that the model is fully compatible with the full data set.

In Figure~\ref{fig:mcmc}, we show the marginalized 1D and 2D PDFs for the model
parameters, and in Table \ref{tab:glo}, we report the values of our best-fit
model parameter values, their medians, and 16--84\% parameter confidence ranges.  
In the left panels of 
Figure~\ref{fig:spec}, we show the stacked best-fit on-source and background spectra for our full
sample and the best-fit global and background models and their residuals.
In this figure, the background data, and corresponding background model, have been rescaled
properly to the source regions, and for illustrative purposes, the data have been
binned to 120 background counts per bin and 12 on-source counts per bin;
however, note that all fits are performed on unbinned data.  The on-source
spectrum is best modeled with significant contributions from the three major
components, including dominant contributions from hot gas at $\simlt$1~keV
({\it red dotted curve\/}), HMXBs at $\simgt$1~keV ({\it blue dashed curve\/}),
and background at $\simlt$0.6~keV and $\simgt$4~keV ({\it green dashed
curve\/}).  


\begin{table}
\renewcommand\thetable{3}
{\footnotesize
\begin{center}
\caption{Global Model Fit Results}
\begin{tabular}{lcr}
\hline\hline
\multicolumn{1}{l}{\sc Parameter} & {\sc Units} & \multicolumn{1}{c}{\sc Best (50\%~$\pm$~34\%)} \\
\hline\hline
                                  $N_{\rm H, gas}$\dotfill &                                                    $10^{22}$~cm$^{-2}$ &      0.56 (0.55$^{+0.12}_{-0.10}$) \\
                                $A_{kT_1}^\dagger$\dotfill &                                                                        &      1.61 (3.00$^{+1.85}_{-1.20}$) \\
                                            $kT_1$\dotfill &                                                                  (keV) &      0.26 (0.24$^{+0.03}_{-0.04}$) \\
                                $A_{kT_2}^\dagger$\dotfill &                                                                        &      0.16 (0.21$^{+0.06}_{-0.09}$) \\
                                            $kT_2$\dotfill &                                                                  (keV) &      0.70 (0.73$^{+0.11}_{-0.11}$) \\
                                $N_{{\rm H,HMXB}}$\dotfill &                                                    $10^{22}$~cm$^{-2}$ &      0.06 (0.50$^{+0.51}_{-0.31}$) \\
                               $A_\Gamma^\ddagger$\dotfill &                                                                        &      0.35 (0.39$^{+0.07}_{-0.05}$) \\
                                          $\Gamma$\dotfill &                                                                        &      1.76 (1.82$^{+0.09}_{-0.10}$) \\
\hline
\multicolumn{3}{c}{Scaling Relations$^\star$} \\
\hline
                      $\log L_{\rm 0.5-8~keV}$/SFR\dotfill &                                            ergs~s$^{-1}$~(\sfr)$^{-1}$ &    40.29 (40.29$^{+0.03}_{-0.03}$) \\
            $\log L_{\rm 0.5-2~keV}^{\rm gas}$/SFR\dotfill &                                            ergs~s$^{-1}$~(\sfr)$^{-1}$ &    39.44 (39.58$^{+0.17}_{-0.28}$) \\
           $\log L_{\rm 0.5-8~keV}^{\rm HMXB}$/SFR\dotfill &                                            ergs~s$^{-1}$~(\sfr)$^{-1}$ &    40.22 (40.19$^{+0.06}_{-0.06}$) \\
                      $\log L_{\rm 0.2-2~keV}$/SFR\dotfill &                                            ergs~s$^{-1}$~(\sfr)$^{-1}$ &    39.96 (39.94$^{+0.04}_{-0.03}$) \\
\hline
\multicolumn{3}{c}{Goodness of Fit Evaluation} \\
\hline
$C$\dotfill &  & 2426\\
$C_{\rm exp}$\dotfill &  & 2576\\
$C_{\rm var}$\dotfill &  & 9908\\
$p_{\rm null}$\dotfill &  & 0.133\\
\hline
\label{tab:glo}
\end{tabular}
\end{center}
$^\dagger$The values of the hot gas normalization represent the quantity $4 \pi d_L^2$/SFR~$10^{-14}/(4 \pi d_A^2 [1+z]^2) \int_V n_{\rm H} n_e dV$~cm$^{-5}$~($M_\odot$~yr$^{-1}$)$^{-1}$~Mpc$^2$, where $d_L$ is the luminosity distance in Mpc, SFR is in units of $M_\odot$~yr$^{-1}$, $d_A$ is the angular diameter distance in cm, $z$ is the redshift, $n_{\rm H}$ and $n_e$ are hydrogen and electron densities in units of cm$^{-3}$.  See https://heasarc.gsfc.nasa.gov/xanadu/xspec/manual/XSmodelApec.html for a full description of the {\ttfamily APEC} model.\\
$^\ddagger$ The power law model normalization has units of photons~keV$^{-1}$~cm$^{-2}$~s$^{-1}$~($M_\odot$~yr$^{-1}$)$^{-1}$~Mpc$^2$ at 1~keV.\\
$^\star$Uncertainties on scaling relations are based on MCMC chains of the scaling relations themselves and represent 16--84\% 1D uncertainties marginalized over all parameters.
}
\end{table}

In the top-right panel of Figure~\ref{fig:spec}, we show the unfolded global
model spectrum, in terms $E L_E$/SFR (where $L_E \equiv \ell_E$~SFR and $E L_E = \nu L_\nu$) and
its HMXB and hot gas model contributions.  The gray shaded region shows the
full range of spectral models for the 68\% of models with the highest posterior probabilities.
The bottom-right panel of Figure~\ref{fig:spec} displays the same spectrum,
but with comparisons to the galaxies presented in \citet{Leh2015} and \citet{Gar2020} (see red stars in Fig.~\ref{fig:samp}
for property comparison).  For ease of comparison, we used gray coloring for
NGC~3310 ({\it dashed\/}) and VV114 ({\it dotted\/}), which have low metallicities
of $Z \approx 0.2 Z_\odot$ and $0.3 Z_\odot$, respectively, that are comparable
to the galaxies in our sample.  The remaining galaxies (i.e., NGC~253,
NGC~3256, and M83, shown as green curves in Fig.~\ref{fig:spec}) are nearly solar metallicity.  

We find that the SFR-normalized \xray\ spectrum for our low-metallicity galaxy sample is
elevated compared to solar-metallicity galaxies, somewhat elevated compared to VV114, and similar
to the low-metallicity galaxy NGC~3310.  The
elevation of our sample \xray\ spectrum over that found in solar-metallicity galaxies
appears to apply to both XRB and hot gas components; however, the effect is of
larger magnitude for the HMXB component.  Integration of our global model
provides a prediction for the luminosity scaling relation with SFR.  Given the
construction of our model, this integration can be performed for both the full
spectrum and portions of the spectrum of interest, such as the HMXB and hot gas
components.  To calculate confidence intervals, we computed such integrations
at each step of our MCMC procedure and thus sampled their marginalized PDFs.
Throughout the remainder of this paper, we choose to assess HMXB scaling relations with
SFR using the 0.5--8~keV band and hot-gas scaling
relations with SFR using the 0.5--2~keV band; hereafter we define these relations as
$\beta_{\rm HMXB}
= L_{\rm 0.5-8~keV}^{\rm HMXB}$/SFR and $\beta_{\rm gas} = L_{\rm 0.5-2~keV}^{\rm gas}$/SFR, where $\beta$ is quoted in units of \lum~(\sfr)$^{-1}$.  These choices were adopted due to
both components providing significant contributions to the overall spectra in the
chosen bands and the availability of published scaling relations that use the
same bandpasses \citep[e.g.,][]{Min2012b,Leh2019,Leh2021}.  

In the upper-right panels of Fig.~\ref{fig:mcmc} we display the 1D and 2D
marginalized PDFs for $\beta_{\rm HMXB}$ and $\beta_{\rm gas}$ and show
comparison values from the local studies of \citet{Leh2019} and \citet{Min2012b}, respectively.  The median values and 16--84\% confidence
intervals for $\beta_{\rm HMXB}$ and $\beta_{\rm gas}$ are 
\begin{equation}
\log L_{\rm 0.5-8~keV}^{\rm HMXB}{\rm /SFR} = \log \beta_{\rm HMXB} =
40.19 \pm 0.06,
\end{equation}
\begin{equation}
\log L_{\rm 0.5-2~keV}^{\rm gas}{\rm /SFR} = \log \beta_{\rm gas} =
39.58^{+0.17}_{-0.28},
\end{equation}
which are listed in Table~\ref{tab:glo} along with the maximum posterior values.  Here,
we note that our derived values of $\beta_{\rm HMXB}$ and $\beta_{\rm gas}$ for
our sample are both significantly elevated compared to local scaling relations.
In \S~\ref{sec:dis} below, we discuss in more detail possible explanations for
the elevation of these relations.

%
%
\begin{figure}
\centerline{
\includegraphics[width=9cm]{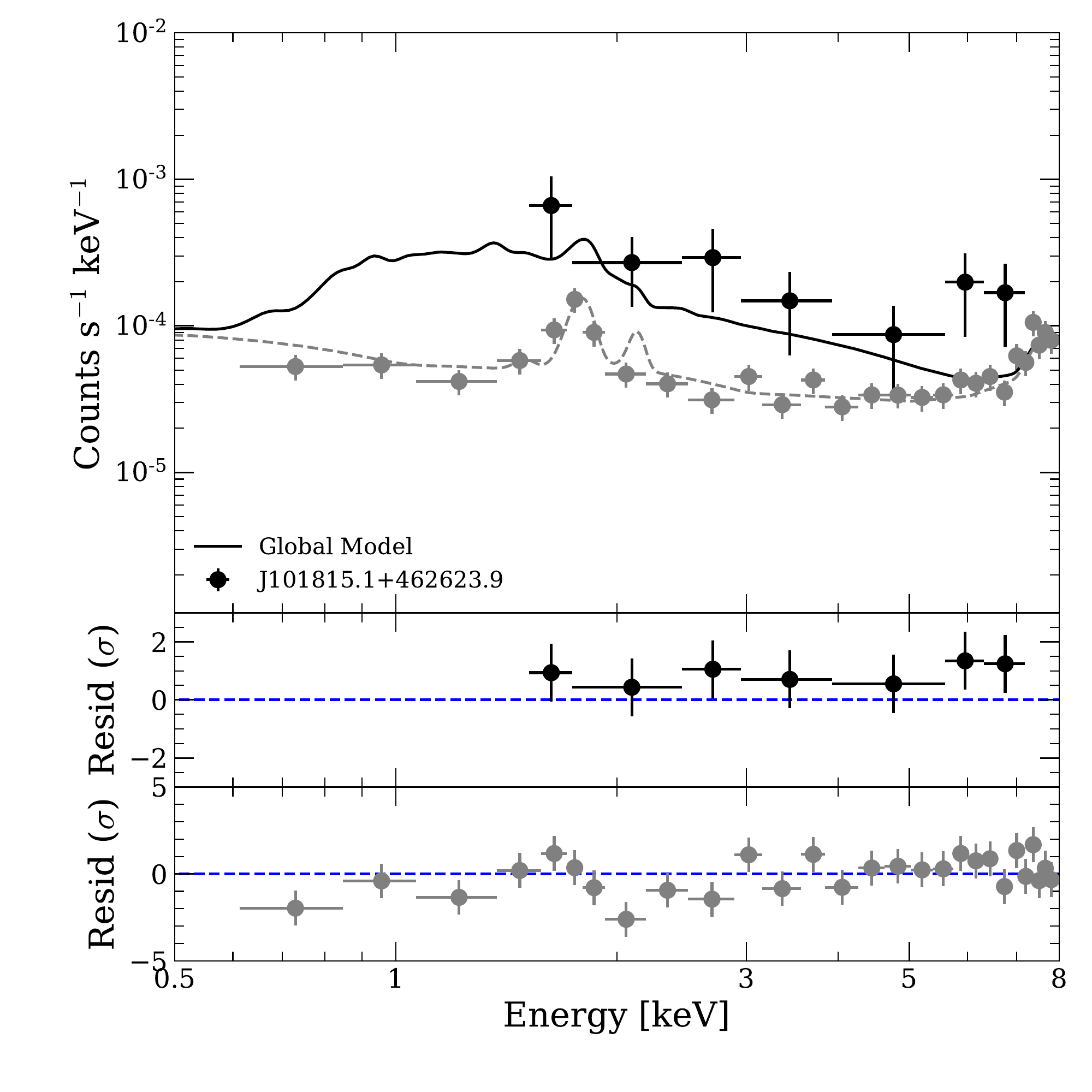}
}
\caption{
\chandra\ spectrum of J101815.1+462623.9 ({\it filled black circle with 1$\sigma$
uncertainties\/}), the only galaxy in our sample that is poorly fit by both our rescaled and
global models ({\it black curve\/}).  The
residuals to the global model ({\it middle panel\/}) suggest that this source has a somewhat flatter \xray\ spectrum than the sample average, and contains notable excesses near the Fe~K region at
6--7~keV.  As such, this object plausibly contains a
heavily-obscured/Compton-thick AGN.
}
\label{fig:out}
\end{figure}

\subsection{Individual Source Models}\label{sub:ind}

To investigate galaxy-to-galaxy variations in the spectra of the galaxies in
our sample, we fit each galaxy spectrum individually using a ``scaled model,''
which consists of our global model (\S~\ref{sub:glo}) rescaled by a
multiplicative constant factor ({\ttfamily CONSTANT} in {\ttfamily xspec}).
Here, all parameters of the global model were held fixed at the global best-fit
values displayed in Table~\ref{tab:glo}, and we fit each galaxy using a single
multiplicative scaling parameter $A_{\rm cnst}$.

For each galaxy, we identified best-fit values and PDFs of $A_{\rm cnst}$ using
$C$ in Eqn~\ref{eqn:cstat} (here we adopt minimum $C$ values as our best-fit models),
and we calculated null-hypothesis probabilities using Eqn~\ref{eqn:pnull}.  In
Col.(6) of Table~2, we report the best-fit values of $A_{\rm cnst}$, the statistical
results of the fit, and calculated 0.5--8~keV fluxes and luminosities for the
single-parameter model (see Col.7--12).  We also provide, in Col.(13)--(16),
the results for the case of the global model (i.e., $A_{\rm cnst} = 1$) for
comparison.  

From random statistical scatter, we expect $\approx$1--2 objects will have
$p_{\rm null}^{\rm src} \simlt 0.05$, which we adopt as a threshold for statistical
acceptability.  We find that all scaled-model fits are statistically
acceptable, with $p_{\rm null}^{\rm src} > 0.05$, suggesting that our modeling does not
require variations in spectral {\it shape} to describe the data.  We note,
however, that our galaxies have small numbers of net counts and poor
constraints on individual bases.  

For the global model itself, which has no free parameters for an individual
galaxy, we find that most galaxies are in good agreement with the direct model
predictions, with the exception of four sources that show some tension with the model
($p_{\rm null}^{\rm gl} \le 0.05$).  Among these four sources, the poorest fitting sources are J165844.5+351923.0 and J101815.1+462623.9.  The former object appears to have a deficit of observed counts, compared to those expected from the relation.  We expect that this could plausibly arise due to stochastic
sampling of the HMXB XLF, which has been shown to produce an additional source
of galaxy-to-galaxy scatter that skews the distribution of $L_{\rm
0.5-8~keV}^{\rm HMXB}$ to low values \citep[see,
e.g.,][]{Gil2004a,Jus2012,Leh2019}.  

The latter galaxy that is poorly fit by our global model,
J101815.1+462623.9, is also the most poorly fit by our scaled model, suggesting
some deviation of the spectral shape of this source with respect to the
global-model shape.  In Fig~\ref{fig:out}, we show the spectrum of
J101815.1+462623.9 and the global model prediction of its spectrum.  Visual
inspection of the spectrum and its residuals to the global model suggest that
this source has a somewhat flatter spectral shape (i.e., lower values of
$\Gamma$) with elevated residuals between 6--7~keV, where the Fe~K line complex
is found.  These spectral features (flat spectral slope and potential Fe~K
feature), along with the fact that our objects were selected to have optical
spectral features consistent with normal star-forming galaxies, suggests that
this object is a good candidate for harboring a heavily-obscured or
Compton-thick AGN.  However, given that this source has been detected with only
$\approx$20 net counts, we do not attempt to derive any detailed parameters
using more complex models.

%
\section{Discussion}\label{sec:dis}
%

\subsection{The HMXB X-ray/SFR Scaling Relation and Scatter}\label{sub:lxs}

\begin{figure}
\centerline{
\includegraphics[width=9cm]{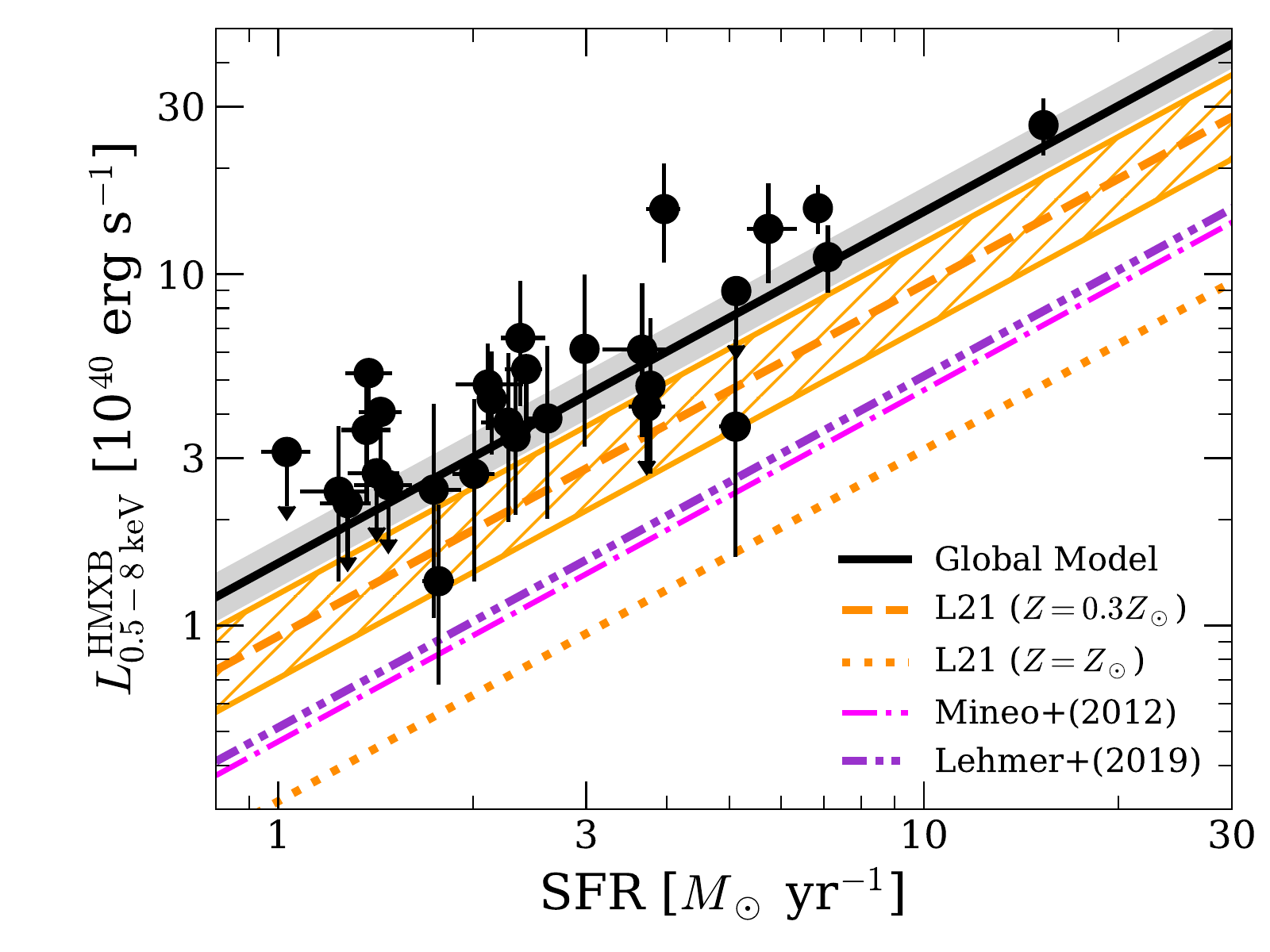}
}
\caption{
HMXB component 0.5--8~keV luminosity, $L_{\rm 0.5-8~keV}^{\rm HMXB}$ versus SFR
for the 30 galaxies in our sample ({\it filled circles with 1$\sigma$ error
bars or 90\% confidence upper limits\/}).  $L_{\rm 0.5-8~keV}^{\rm HMXB}$ was
calculated for each galaxy using the scaled model described in
\S~\ref{sub:ind}.  Our global best-fit model scaling relation, $\beta_{\rm
HMXB}$, is displayed as a solid black line with a grey shaded region
representing the 1$\sigma$ uncertainty.  For comparison, we show the local HMXB
relations from \citet[][{\it magenta dot-dashed line}]{Min2012a} and
\citet[][{\it purple dot-dot-dashed line\/}]{Leh2019}, as well as the
metallicity-dependent scaling relation from \citetalias{Leh2021} for solar
metallicity ({\it dotted orange line\/}) and 0.3~$Z_\odot$ ({\it dashed orange
line with 1$\sigma$ hatched region\/}), the latter of which is representative of the metallicity of our
sample.
}
\label{fig:lxs}
\end{figure}

In Figure~\ref{fig:lxs}, we display the HMXB model component luminosity,
$L_{\rm 0.5-8~keV}^{\rm HMXB}$, versus SFR for our sample.  The HMXB
luminosities were computed using the scaled model fits to each galaxy.  We have
overlaid our global scaling relation, $\log \beta_{\rm HMXB} = 40.19 \pm 0.06$.  Given that the global model
provides a reasonable description of the majority of the galaxy \xray\ spectra without
adjustment, it is not surprising that the global model agrees well with the
majority of the $L_{\rm 0.5-8~keV}^{\rm HMXB}$ values.  

The metallicity-dependent HMXB XLF work by \citetalias{Leh2021} provides direct
constraints on the $L_{\rm 0.5-8~keV}^{\rm HMXB}$-SFR relation as a function of
metallicity through the integration of their HMXB XLF models.  By construction,
our sample contains a narrow distribution of metallicities, with mean and
1$\sigma$ standard deviation of \lgoh~=~$8.16 \pm 0.03$ ($Z = [0.29 \pm
0.02$]~$Z_\odot$).  At the sample mean, \citetalias{Leh2021} predict $\log
\beta_{\rm HMXB}^{\rm L21}(Z=0.3 Z_\odot) = 39.97 \pm 0.12$, which we display
in Figure~\ref{fig:lxs} as a dashed orange line with hatched uncertainty region.  This value is somewhat lower than the value found for our sample, albeit within the current uncertainties. 

%
%

For comparison, we also display Figure~\ref{fig:lxs}, the solar-metallicity
model from \citetalias{Leh2021}, which lies at $\log \beta_{\rm HMXB}^{\rm
L21}(Z=Z_\odot) = 39.50 \pm 0.11$, a factor of $\approx$4 times lower than our
sample.  We also show ``local'' estimates of $\log \beta_{\rm HMXB}$ from the
work of \citet{Min2012a} and \citet{Leh2019}, which are based on samples of
nearby star-forming galaxies that span a non-negligible range of metallicity.
After correcting for differences in the \citet{Min2012a} IMF, these samples
have values of $\log \beta_{\rm HMXB} = 39.67 \pm 0.06$ and $\log \beta_{\rm
HMXB} = 39.71^{+0.14}_{-0.09}$ for \citet{Min2012a} and \citet{Leh2019},
respectively.  These values lie between those of our sample and the
\citetalias{Leh2021} solar-metallicity prediction.  For the \citet{Leh2019}
sample, which quotes metallicity values, galaxies with sSFR~$\simgt
10^{-10}$~yr$^{-1}$, which are dominated by HMXB populations, have
metallicities of $\approx$0.8~$Z_\odot$.  At this metallicity,
\citetalias{Leh2021} predict $\log \beta_{\rm HMXB}^{\rm L21}(Z=0.8 Z_\odot) =
39.64 \pm 0.11$, consistent with the \citet{Leh2019} value for local galaxies.

The above comparisons of point-source studies of HMXBs with the constraints
from our galaxies, suggests near consistency when the $L_{\rm
X}$(HMXB)/SFR scaling relation is put into context with galaxy metallicity.  To
further test this connection, we investigated the scatter of the \xray\
emission from our galaxy population, as measured by the distribution of
0.5--8~keV counts.  As discussed in \S~5.1 of \citetalias{Leh2021}, incomplete
sampling of the HMXB XLF can lead to non-negligible scatter of $L_{\rm
X}$(HMXB) for a given SFR.  The magnitude of this scatter is predicted to
increase with decreasing SFR.  For the \citetalias{Leh2021} HMXB XLF model at
$\approx$0.3~$Z_\odot$, the scatter is expected to decline from
$\approx$0.5~dex to 0.1~dex over SFR~=~1--10~\sfr, a range covered by our
galaxy sample.

%
%
\begin{figure}
\centerline{
\includegraphics[width=9cm]{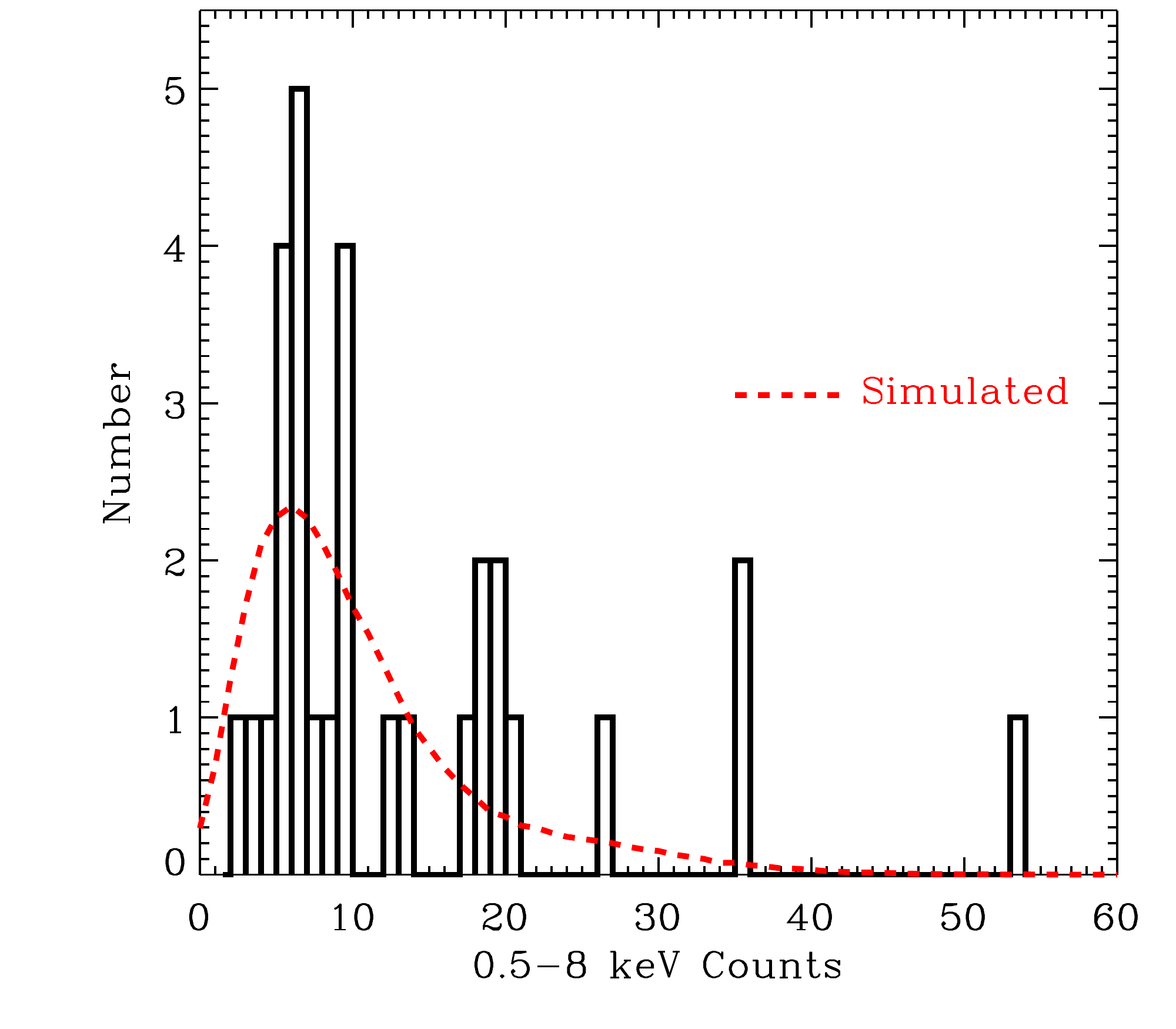}
}
\caption{
Observed distribution of on-source 0.5--8~keV counts ({\it black histogram};
see Col.(4) in Table~\ref{tab:fit}).  The dashed red curve shows the expected
distribution from simulations that include HMXB emission and stochastic
fluctuations from sampling of the \citetalias{Leh2021} HMXB model XLF, plus minor
contributions from hot gas, local background levels, and Poisson fluctuations.
A K-S test indicates that the observed and expected distributions are
consistent, suggesting the scatter in our galaxy sample is also consistent with
expected scatter from HMXB populations.
}
\label{fig:scat}
\end{figure}

To test more explicitly whether our sample data are consistent with HMXB XLF model framework from \citetalias{Leh2021}, we simulated the expected distributions of on-source counts for each of our galaxies and compared those distributions to our observations.  For a given galaxy, we
first used the SFR and \lgoh\ values as input to the \citetalias{Leh2021} HMXB
XLF model, which specifies the expected HMXB XLF shape and normalization for
the galaxy.  Next, treating the HMXB XLF model as a probability distribution
function, we drew HMXB luminosities from that distribution to generate
simulated HMXB populations that could plausibly be expected from within the
galaxy.  Summing the luminosity contributions from a given simulated HMXB
population provides an estimate of the integrated HMXB population luminosity.
We performed 5000 simulations for each galaxy to form a distribution of
expected HMXB luminosities expected from the population.  We then converted the
simulated HMXB luminosities into contributions to the source counts, using the
$L_{\rm 0.5-8~keV}^{\rm HMXB}$--to--counts conversion factor appropriate for
the galaxy.  We then added the background count estimate from Col.(5) in
Table~\ref{tab:fit} and the expected hot gas model contribution to the total
counts based on our best-fit model.  We note that while some variation in the
intrinsic hot gas emission is expected, this variation has not been well
characterized observationally or theoretically and is not included in our
simulations.  However, our best-fit model predicts that HMXBs provide a factor of
$\approx$2--4 (median 3.8) times more counts than the hot gas component,
suggesting that unmodeled variations in hot gas emission are likely to have a
negligible impact on our results.  

For a given simulation, the combined HMXB, hot gas, and background counts
estimates for a given galaxy are summed and subsequently perturbed using a
Poisson distribution with a mean equal to the summed counts.  This results in
5,000 distributions of simulated on-source counts for the sample.  Combining
all of the simulations together provides a smooth distribution of
model-expected on-source counts for our sample, which we display in
Figure~\ref{fig:scat}, along with the actual observed counts distribution.  A
two-sided K-S test between the simulated counts and our data suggest that the
two distributions are statistically consistent with each other ($p_{KS} = 0.18$).  Thus, the
observed distribution of counts from our sources is consistent with the
combination of noise from our data and expected HMXB XLF scatter.

%
%
\begin{figure*}
\centerline{
\includegraphics[width=16cm]{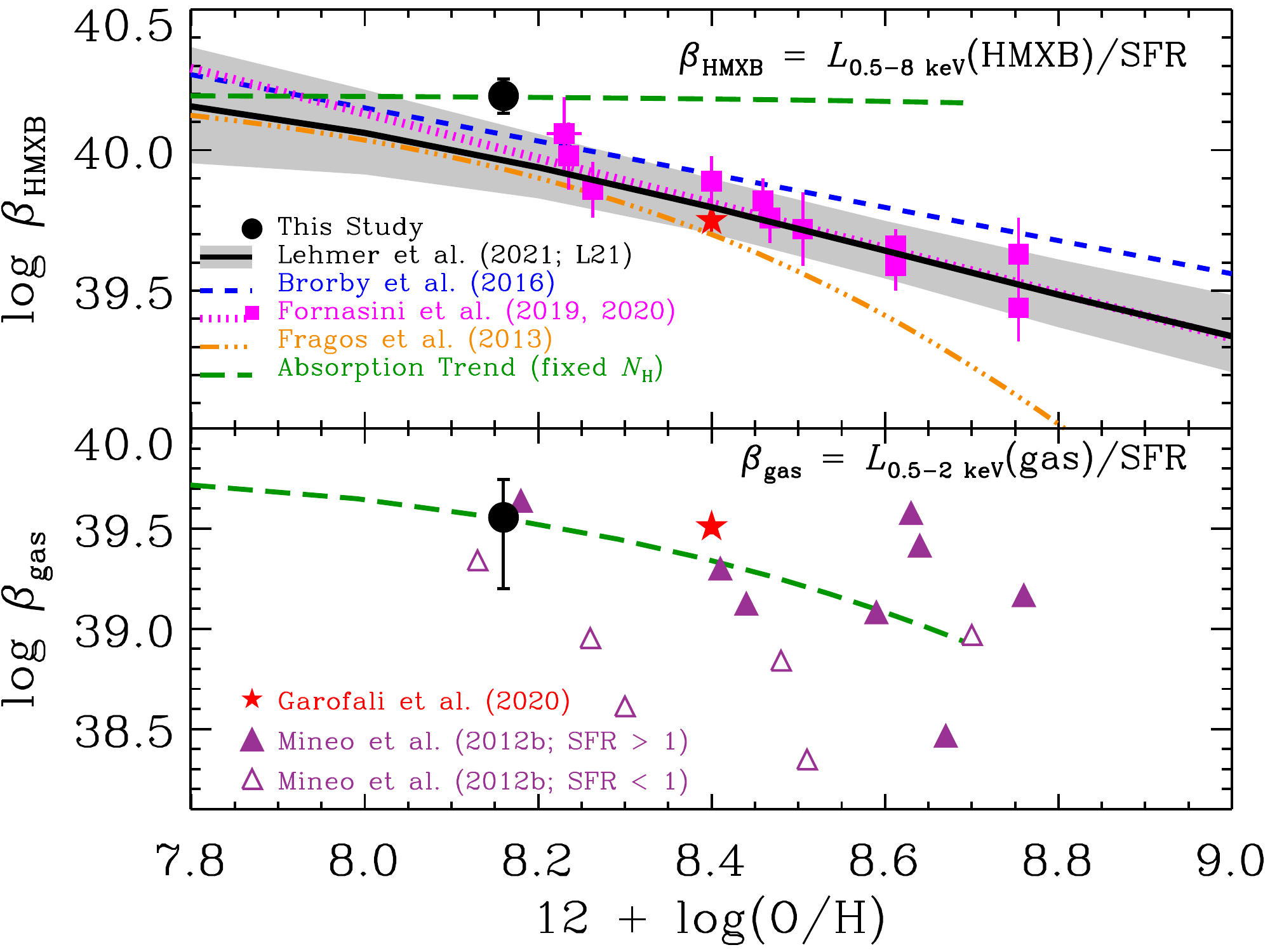}
}
\caption{
($a$) Logarithm of the 0.5--8~keV HMXB luminosity per SFR, $\beta_{\rm HMXB}$
versus galaxy metallicity for our galaxy sample ({\it black filled circle with
1$\sigma$ uncertainties\/}) and 12 stacked subsamples of $z \approx$~0.1--2.6
galaxies from \citet[][{\it magenta filled squares with 1$\sigma$
uncertainties\/}]{For2019,For2020}.  The \citetalias{Leh2021} relation and its
16--84\% confidence interval are shown as a solid curve and gray shaded 
region, respectively.  For comparison, the empirical relations from \citet[][{\it dotted magenta
curve\/}]{For2020} and \citet[][{\it short-dashed blue curve\/}]{Bro2016},
as well as the XRB population synthesis model prediction from \citet[][{\it
triple-dot-dashed orange curve}]{Fra2013a} are overlaid.  Our new constraint
on $\beta_{\rm HMXB}$ is somewhat elevated compared to that expected from $L_{\rm X}$-SFR-$Z$ relations in the literature, potentially due to our galaxies containing relatively young and unobscured HMXB populations.  The {\it green long-dashed} curve highlights the
expected effect of varying metallicity for a fixed column density of $N_{\rm H} = 6 \times 10^{21}$~cm$^{-2}$, and
illustrates that the data are not consistent with being driven solely by the
effects of metallicity on absorption.
($b$) Logarithm of the 0.5--2~keV hot gas luminosity per SFR, $\beta_{\rm
gas}$, versus metallicity (\lgoh) for our sample and the point-source-excised
analyses of 16 galaxies with metallicity estimates in \citet[][{\it purple triangles\/}]{Min2012b} and VV114 \citep[][{\it filled
red star\/}]{Gar2020}.  The \citet{Min2012b} sample spans a broader range of
SFR than our sample, and we use {\it filled triangles} to represent galaxies
with SFR~$> 1$~\sfr, consistent with the range of our sample, and {\it open
triangles} represent galaxies with SFR~$< 1$~\sfr.  The effects of metallicity
on absorption ({\it green long-dashed curve\/}) are expected to be significant
for $\beta_{\rm gas}$ and may play a role in the observed trend.
}
\label{fig:met}
\end{figure*}

\subsection{Metallicity Dependence of HMXB and Hot Gas Scaling Relations}\label{sub:met}

In \S~\ref{sub:glo}, we showed that our global fit yields constraints on the
scaling relations $\beta_{\rm HMXB}$ and $\beta_{\rm gas}$ that are elevated
compared to scaling relations derived for more representative local galaxies
\citep[e.g.,][see marginalized distributions in
Fig.~\ref{fig:mcmc}]{Min2012a,Min2012b,Leh2019}.  As shown in
\S\ref{sub:lxs}, the elevation of $\beta_{\rm HMXB}$ can be attributed primarly to the
metallicity-dependence of the $L_{\rm X}$(HMXB)-SFR-$Z$ relation, as has been presented in the literature (see \S\ref{sec:intro}); however, there is some evidence that our relation is further enhanced over such relations.  

To provide context to our results in terms of potential metallicity
dependencies, we constructed Figure~\ref{fig:met}, which displays $\beta_{\rm
HMXB}$ (Fig.~\ref{fig:met}$a$) and $\beta_{\rm gas}$ (Fig.~\ref{fig:met}$b$) as
a function of metallicity. 
The black curve in Figure~\ref{fig:met}$a$ shows the \citetalias{Leh2021}
relation and its 1$\sigma$ uncertainty.  
For
comparison, we also show the XRB population synthesis prediction from \citet[][{\it orange triple-dot-dashed curve\/}]{Fra2013a}, the $L_{\rm X}$(HMXB)-SFR-$Z$ relations from
\citet[][{\it blue short-dashed line}]{Bro2016} and \citet[][{\it magenta
dotted line\/}]{For2020}, and the stacked constraints from \citet[][{\it magenta
filled squares with 1$\sigma$ error bars\/}]{For2019,For2020}, the latter of which are based
on $z \approx$~0.1--2.6 galaxy samples that include 20--400 galaxies per data point.  
In this context, our estimate of $\beta_{\rm HMXB}$ ({\it black
filled circle with 1$\sigma$ error bars\/}) is elevated by a factor of 1.2--1.5 times that expected from
\citetalias{Leh2021}, \citet{For2020}, and \citet{Bro2016} relations.  

We speculate that an elevated value of $\beta_{\rm HMXB}$ for our sample, compared with $L_{\rm X}$(HMXB)-SFR-$Z$ relations may be expected due to our sample having (1) somewhat lower HMXB intrinsic absorption than other samples (e.g., $N_{\rm H,HMXB} \simlt 10^{21}$~cm$^{-2}$ for our sample versus $\approx 6 \times 10^{21}$~cm$^{-2}$ from \citetalias{Leh2021}; see $\S$~\ref{sub:glo}); and (2) relatively high sSFRs, and thus, younger stellar populations compared to the galaxies used to derive local relations (see Fig.~\ref{fig:samp} for comparison with other samples).
Regarding the latter point, a recent study of galaxies in the \chandra\ Deep Field-South by \citet{Gil2021} found that $L_{\rm X}/M_\star$ for HMXBs declines by nearly an order of magnitude from 10~Myr to 100~Myr.  As such, galaxies with relatively young stellar populations, like our sample, may have elevated $\beta_{\rm HMXB}$ compared to more representative galaxy samples.  Additional evidence for enhanced \xray\ emission (relative to relations) has been noted for galaxy samples that are selected explicitly to have signatures of very young stellar populations \citep[e.g., from Lyman-continuum emitters and green peas;][]{Ble2019,Svo2019,Fra2022}.  For example, \citet{Svo2019} find that the two \xray-detected green peas in their sample have soft spectra with \xray\ luminosities on the order of $10^{42}$~\lum.  Given their SFR~$\approx$~20--40~\sfr\ (after correcting to our adopted IMF; see green downward triangles in Fig.~\ref{fig:samp}), these $L_{\rm X}$/SFR values are $\approx$2--3 times higher than observed for the galaxies in our sample.  Considering the stellar masses of these galaxies of $\approx$few~$\times 10^{9}$~\msol, this implies $L_{\rm X}/M_\star \approx 10^{31}$~ergs~s$^{-1}$~$M_\odot^{-1}$, a value consistent with stellar populations of $\simlt$100~Myr in age \citep[see][for details]{Gil2021}.

Fewer constraints on the metallicity dependence of $\beta_{\rm gas}$ are
available in the literature, and no formal relations have been proposed.  In
Figure~\ref{fig:met}$b$, we show our estimate of $\beta_{\rm gas}$ along with
estimates for 14 individual galaxies from \citet[][{\it purple
triangles\/}]{Min2012b} that have metallicities extracted from the literature (see \citealt{Bas2013b} for details) and VV114 from \citet[][{\it
red star\/}]{Gar2020}.  The \citet{Min2012b} and \citet{Gar2020} constraints on
$\beta_{\rm gas}$ are based on careful measurements of the diffuse emission after
excising \xray\ point source contributions and are expected to be highly
reliable.  A Spearman's rank test suggests that there is no significant
correlation between $\beta_{\rm gas}$ and \lgoh, when including our constraint
and those in the literature ($\rho = -0.32$ for 16 data points).  However, if
we restrict the comparison sample to include only galaxies with
SFR~$\simgt$~1~\sfr\ that are comparable to our sample and less subject to
statistical scatter than lower-SFR galaxies, we find a 95\% significant ($\rho
= -0.63$ for 10 data points) anticorrelation.

If $\beta_{\rm gas}$ is indeed anticorrelated with metallicity, there are
potential physical reasons that could explain such a trend.  As highlighted in
\S\ref{sub:fit}, hot gas emission has been studied extensively in nearby
galaxies (see discussion and citations in \S\ref{sub:fit}).  While we expect
that our simple thermal model -- a two-temperature plasma with absorption by an
ISM with solar abundance -- will be sufficient to extract a reliable
measurement of $\beta_{\rm gas}$, we do not assume that our model provides a
faithful description of the full physical picture.  Detailed studies of
resolved nearby galaxies find that distributions of plasma temperatures are
inevitably present \citep[e.g.,][]{Str2000,Str2004,Kun2010,Lop2020,Wan2021} and the
efficiency of converting mechanical heating of the ISM into hot gas \xray\
emission depends on star-formation timescales that are shorter than those
measured for typical galaxies \citep[e.g.,][]{McQ2018,Gil2021}.  Therefore, it
seems plausible that variations in star-formation history and physical
environment (e.g., galaxy morphology and gravitational potential) will lead to
variations in $\beta_{\rm gas}$.  For local galaxies, these factors are often
correlated with metallicity.  We can also expect that both the absorbing and
emitting ISM will be influenced by metallicity.  For a fixed hydrogen column
density and fixed {\it intrinsic} $L_{\rm X}^{\rm gas}$/SFR ratio, the
escaping low-energy emission (i.e., $\beta_{\rm gas}$) will decline with
increasing metallicity due to the increasing impact of metal absorption
lines (particularly from C, O, Ne, and Fe L).  

While the low signal-to-noise \xray\ spectra in this study and in the literature
are insufficient to reliably constrain directly the metallicities of the
absorbing ISMs, we can determine theoretically the effect of varying ISM abundance on
emergent \xray\ emission.  In Figure~\ref{fig:met}, we show how $\beta_{\rm
HMXB}$ and $\beta_{\rm gas}$ would be impacted by metallicity variations for
fixed values of $N_{\rm H}$ and intrinsic $L_{\rm X}$/SFR ({\it green dashed
curves\/}).  The displayed curves are anchored to the best-fit values of
$\beta_{\rm HMXB}$ and $\beta_{\rm gas}$ and mean \lgoh\ from this study and assume the
best-fit $N_{\rm H}$ values (see Table~\ref{tab:glo}).  Since absorption
primarily affects the emergent low-energy emission, $\beta_{\rm HMXB}$
(calculated for the 0.5--8~keV band) is not strongly impacted by
metallicity-dependent absorption, and the observed $L_{\rm X}$(HMXB)-SFR-$Z$
relation is inconsistent with being driven by absorption (at least in the
0.5--8~keV band).  However, the predicted impact on $\beta_{\rm gas}$ (derived
in the 0.5--2~keV band) is significant across $Z \approx$~0.2--1~$Z_\odot$, and
the metallicity-dependent trajectory appears to be consistent with the
published constraints for SFR~$\simgt$~1~\sfr\ galaxies (by visual inspection).  

While a detailed investigation of the dependencies of $\beta_{\rm gas}$ on
physical properties is beyond the scope of the current paper, this tantalizing
result calls for future studies on how hot gas emission varies with galaxy
properties like metallicity and star-formation history.  As we outline in the
next section, since hot gas emission dominates at low \xray\ energies in
star-forming galaxies, it may also provide important contributions to ISM
ionization and IGM heating in low-metallicity galaxies.

\subsection{The X-ray--to--IR Emergent and Intrinsic SED}\label{sub:ext}

As discussed in \S\ref{sec:intro}, sources of high-energy emission, like hot
gas and HMXBs, could provide substantial long-range heating of the IGM in the early Universe and 
ionizing radiation to galaxy ISMs.
However, the potential
for these sources to have significant impacts depends on how these models are extrapolated
into the EUV and soft \xray\ bands (0.01--0.5~keV), where no direct observational constraints are available.  Here we
provide model constraints on the SFR-normalized SED of our sample
across a broad wavelength range, spanning the near-IR ($\approx$1$\mu$m) to hard
X-rays ($\approx$30~keV).  Our models include contributions from stellar,
nebular, and dust models from {\ttfamily Lightning}, as discussed in
\S\ref{sec:phys}, and global-model constraints to the hot gas and HMXB
emission (see \S\S\ref{sub:fit} and \ref{sub:glo}).


\begin{deluxetable*}{@{\extracolsep{4pt}}cccccccccc@{}}
\renewcommand\thetable{4}
\tablewidth{1.0\columnwidth}
\tabletypesize{\footnotesize}
\tablecaption{SFR-Normalized Model SED}
\tablehead{
\multicolumn{2}{c}{}  &  \multicolumn{8}{c}{$EL_E$/SFR~($10^{40}$~\lxsfru)} \\
\multicolumn{2}{c}{}  &  \multicolumn{4}{c}{Emergent SED} & \multicolumn{4}{c}{Intrinsic SED} \\
\cline{3-6}
\cline{7-10}
%
%
\multicolumn{1}{c}{$\log \lambda($\AA$)$} &  \multicolumn{1}{c}{$\log E({\rm keV})$} & \colhead{stellar} & \colhead{hot gas} & \colhead{HMXBs} & \colhead{Total} & \colhead{stellar} & \colhead{hot gas} & \colhead{HMXBs} & \colhead{Total} \\
\vspace{-0.25in} \\
\multicolumn{1}{c}{(1)} & \multicolumn{1}{c}{(2)} & \colhead{(3)} & \colhead{(4)} & \colhead{(5)} & \colhead{(6)} & \colhead{(7)} & \colhead{(8)} & \colhead{(9)} & \colhead{(10)}
}
\startdata
4.07 & $-$2.98 & 441 & 0.0150 & 0.0552 & 442 & 445 & 0.0150 & 0.0552 & 446 \\
4.02 & $-$2.93 & 493 & 0.0084 & 0.0569 & 494 & 500 & 0.0084 & 0.0569 & 501 \\
3.98 & $-$2.88 & 531 & 0.0000 & 0.0587 & 532 & 542 & 0.0000 & 0.0587 & 543 \\
3.93 & $-$2.84 & 554 & 0.0110 & 0.0605 & 555 & 569 & 0.0110 & 0.0605 & 569 \\
3.89 & $-$2.79 & 592 & 0.0462 & 0.0624 & 592 & 611 & 0.0462 & 0.0624 & 612 \\
3.84 & $-$2.75 & 650 & 0.0232 & 0.0644 & 650 & 668 & 0.0232 & 0.0644 & 669 \\
3.79 & $-$2.70 & 698 & 0.0173 & 0.0665 & 698 & 733 & 0.0173 & 0.0665 & 733 \\
3.75 & $-$2.65 & 752 & 0.0866 & 0.0687 & 753 & 800 & 0.0866 & 0.0687 & 800 \\
3.70 & $-$2.61 & 848 & 0.0184 & 0.0709 & 849 & 882 & 0.0184 & 0.0709 & 883 \\
3.66 & $-$2.56 & 878 & 0.0305 & 0.0737 & 879 & 959 & 0.0305 & 0.0737 & 960 \\
\enddata
\tablecomments{
This table is available in its entirety in machine-readable form.  Only an abbreviated version of the table is shown here to illustrate form and content.
}
\label{tab:mod}
\end{deluxetable*}

For the purpose of obtaining interpolations of our models into the EUV and soft \xray\
range that are as realistic as possible, we modified the stellar and HMXB
components used to model our data in the following ways.  For the stellar model, we utilized our
{\ttfamily Lightning}-based SED for wavelengths shorter than the Lyman break at
912~\AA, and extrapolated the models into the EUV using {\ttfamily BPASS}
v.~2.2.1 \citep{Eld2017,Sta2018} SEDs.  This choice is motivated by the fact that
the {\ttfamily BPASS} models both extend further into the EUV than those of
{\ttfamily Lightning} and include modeling of stellar atmospheres in
interacting binary stars, which can provide important contributions to the EUV.
For this extrapolation, we adopted the binary-star BPASS models corresponding
to a metallicity of $Z \approx 0.3 Z_\odot$ and a \citet{Cha2003} IMF with an
upper-mass cutoff at 300~$M_\odot$.  When adopting the SFH obtained by
{\ttfamily Lightning}, we found that the {\ttfamily BPASS} SED was similar to
our {\ttfamily Lightning}-based SED at wavelengths longer than the Lyman break,
where the SEDs are constrained by our data.

We modified the HMXB model component from a simple power-law model to a more
physically motivated ULX model that is consistent with our \xray\ data.  To
identify a suitable physically motivated HMXB component, we inspected the ULX
spectra studied by \citet{Wal2018} and found that the spectral model for Ho~IX X-1 was the
most similar in 0.5--8~keV shape to that of our \xray\ spectral model and chose to adopt
the Ho~IX X-1 model fits for our extrapolations.  \citet{Wal2018} model Ho~IX
X-1 using the combination of a standard radiatively-efficient accretion disk
({\ttfamily DISKBB}) for the outer portions of the accretion disk, a
geometrically thick disk with modified temperature gradient ({\ttfamily
DISKPBB}) for the inner portion of the accretion disk, and a cut-off power-law
to account for Comptonization from either an accretion column (for a NS
accretor) or a funnel-like beaming medium \citep[BH case][]{Wal2017}.  We thus
constructed our modified HMXB SED component using the Ho~IX X-1 model from
\citet{Wal2018}, but with our best-fit value of the absorption column density
and normalization adjusted to fit our data.  We found that this approach
produced a nearly equivalent quality fit to the observed data as that reported in
\S\ref{sub:glo}.

In Figure~\ref{fig:ext}, we show the resulting model over the energy range $E
=$~0.001--30~keV ($\lambda =$~0.4--12,400\AA) both as the emergent SED
(Fig.~\ref{fig:ext}$a$) and intrinsic SED (Figs.~\ref{fig:ext}$b$ and
\ref{fig:ext}$c$) models.  The emergent SED model includes stellar, nebular,
and dust emission, attenuation from the nebula and dust, as well as attenuated
hot gas and HMXB emission.  The intrinsic SED model includes unattenuated
stellar, hot gas, and HMXB emission, and does not include attenuation and
emission from nebulae and dust.  In Table~\ref{tab:mod}, we tabulate the numerical values
of our model spanning the broader wavelength range of 1$\mu$m to 30~keV.  We
also provide model-based uncertainties for our emergent SED following the
procedures discussed in $\S$\ref{sub:glo}, and the gray shaded region in
Figure~\ref{fig:ext}$a$ represents 16--84\% confidence intervals for both the
combined {\ttfamily Lightning} SED models and our global model fits to the \chandra\
data.  Despite our efforts to provide uncertainties, we emphasize that these
uncertainties are \emph{underestimated} in the EUV range (\hbox{0.01--0.5~keV}), as
they only include uncertainties on extrapolated model components constrained
outside of this range.  Additional emission in the EUV range may arise from
varied model extrapolations or additional hot thermal components that do not
contribute significantly outside the EUV.

\subsubsection{The Emergent SED}

\begin{figure*} 
\centerline{ \includegraphics[width=18cm]{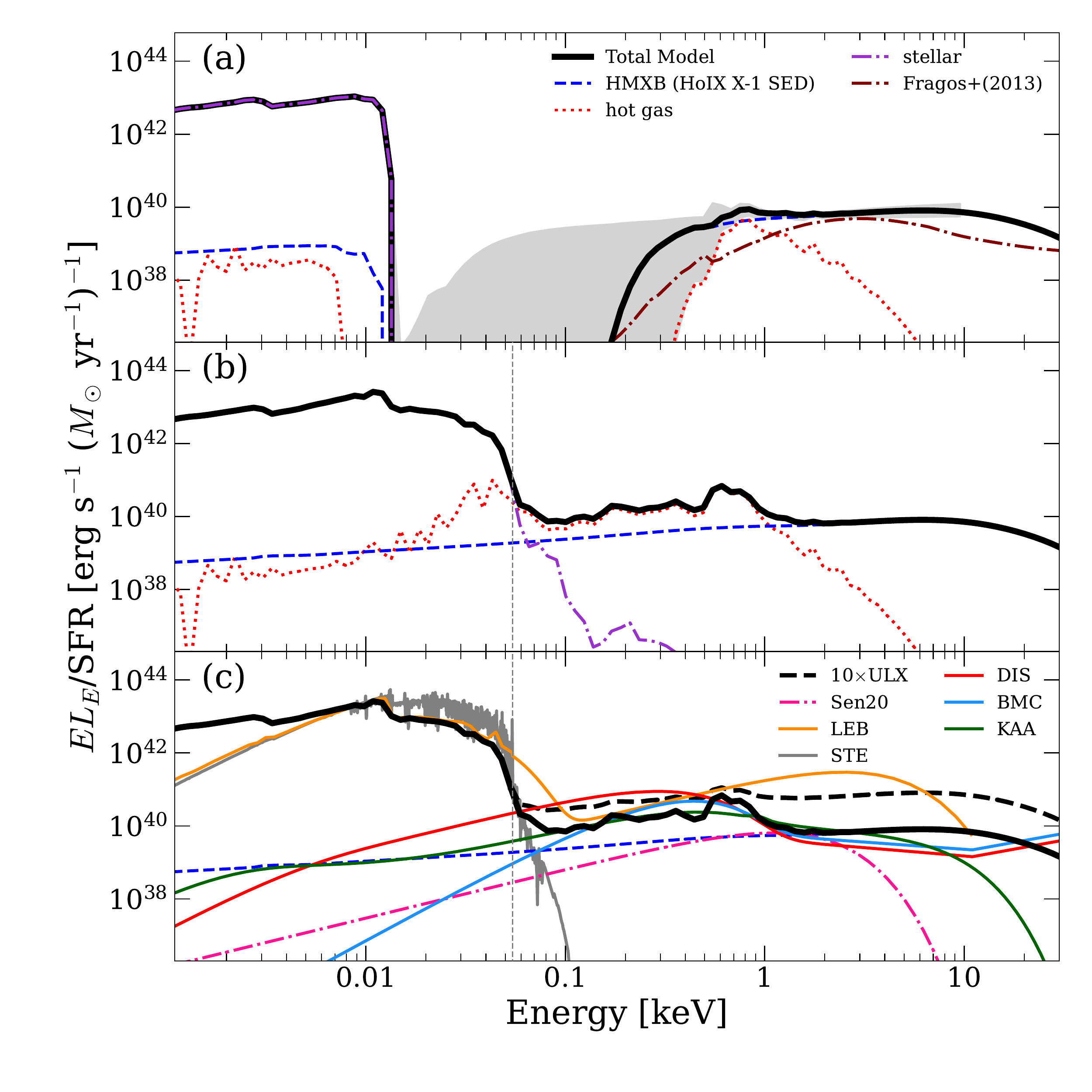} }
\vspace{-0.2in}
\caption{
Broad band ($E=$~0.001--30~keV; $\lambda =$~0.4--12,400~\AA) SFR-normalized
model SEDs for our full sample.  We display both the ``observed'' ($a$) and
``intrinsic'' ($b$ and $c$) SED models.  The observed SED model consists of
contributions from stellar, nebular, and dust emission ({\it dot-dashed
lavendar curve\/}), along with attenuated emission from hot gas ({\it dotted
red curve\/}) and HMXBs ({\it dashed blue curve\/}) (see \S~\ref{sub:ext} for
details).  The gray shaded region surrounding the total observed model
represents the 16--84\% confidence range for the best-fit models.  We note that
these uncertainties are expected to be underestimated in the EUV
(0.01--0.5~keV), where no data are present.  For comparison, we show that the
\citet{Fra2013a} XRB models for $Z \approx 0.3 Z_\odot$ galaxies ({\it
dot-dashed maroon curve\/}) compare well with our HMXB SED constraint.  The
\citet{Fra2013a} model has been used in assessing the impact of \xray\ emission
on the IGM in the high-redshift Universe \citep[e.g.,][]{Das2017}.
The intrinsic SED model consists of the stellar emission, without
nebular or dust absorption and emission, plus the unattenuated hot gas
and HMXB emission (see panel $b$).  
In panel $c$, we show the intrinsic model compared with other SEDs in the
literature.  These include, the model for XRBs from \citet{Sen2020}
(Sen20) at $L_{\rm X}$/SFR~$= 10^{40}$~\lum, and various SEDs presented in
\citet{Sim2021} (see annotations and discussion in \S~\ref{sub:sub:int} for
details).  Statistical fluctuations
of the ULX populations in dwarf galaxies are common and could lead to elevated
$L_{\rm X}$/SFR values (e.g., the {\it dashed black curve\/} representing a
factor of 10 enhanced ULX rate) that could yield high levels of ionizing
radiation sufficient to power He~\small{II} emission.
} 
\label{fig:ext} \end{figure*}

The emergent SED of our sample (Fig.~\ref{fig:ext}$a$) provides a benchmark for
comparing with 21~cm studies of the impact of \xray\ heating on the
high-redshift IGM (see \S\ref{sec:intro} for discussion).  A recent first
result from the HERA collaboration \citep{Her2021} provided new upper limits on
the 21~cm power-spectrum fluctuations from the IGM at $z \approx$~8 and 10.
Using the HERA limits, along with galaxy and IGM property constraints (e.g.,
galaxy UV luminosity function and Ly$\alpha$ forest constraints on the IGM
opacity), they placed constraints on the spin temperature and the average
galaxy $L_{\rm <2~keV}^{\rm ion}$/SFR ratio required to heat the IGM to those
levels.  The quantity $L_{\rm <2~keV}^{\rm ion}$/SFR is defined as the ionizing
\xray\ radiation between 0.2--2~keV that escapes to the IGM.  

As discussed in \S\ref{sec:samp}, our galaxy sample was selected to have
properties that were comparable to star-forming active
galaxies that may have provided a substantial fraction of the \xray\ radiation
field at $z \sim$~6--10.  Although a detailed estimate of the average $L_{\rm
<2~keV}^{\rm ion}$/SFR ratio for galaxies at $z \sim 10$ would require detailed
knowledge of the distribution of metallicities and \xray\ spectral models for
galaxies with a variety of properties, it is instructive to compare our sample
constraints with those from HERA.  For our models, we calculate the quantity
$L_{\rm 0.2-2~keV}$/SFR (equivalent to $L_{\rm <2~keV}^{\rm ion}$/SFR) and
propagate its uncertainties following the methods discussed in \S\ref{sub:glo}.
The HERA collaboration find $\log L_{\rm <2~keV}^{\rm ion}$/SFR~=~40.2--41.9
(68\% confidence highest posterior density), as compared with our value of
$\log L_{\rm 0.2-2~keV}$/SFR~=~$39.94^{+0.04}_{-0.03}$ for our model.  We note that
the HERA 1D PDF has a broad tail toward low values of $\log L_{\rm
<2~keV}^{\rm ion}$/SFR, and our value of $\log L_{\rm 0.2-2~keV}$/SFR is
well within the 95\% confidence range, which extends to just below $\log L_{\rm
0.2-2~keV}$/SFR~$\approx 39$.  Thus, our constraints are currently
compatible with those of HERA.  Future, improved constraints from HERA are
expected to help put into context the \xray\ spectral constraints established here
and the \xray\ radiation field generated from galaxies in the early Universe
(at $z \approx 10$).

%
%

\subsubsection{The Intrinsic SED}\label{sub:sub:int}

Turning now to the intrinsic SED, the EUV and soft-\xray\ components of the
spectrum provide a measure of the ionizing photon budget for a variety of
atomic species, including those that have been observed in nebulae, but are not
readily produced by stellar populations (e.g., He~{\small II}, C~{\small IV},
O~{\small IV}, and Ne~{\small V}).  As discussed in \S\ref{sec:intro},
He~{\small II} nebular emission, in particular, has been studied extensively in
the literature, and its connection to \xray\ emitting sources is under debate.
For example, \citet{Sch2019} showed that the intensity of the high-ionization
He~{\small II} $\lambda$4686 line, relative to H$\beta$ $\lambda$4861,
correlates with metallicity, following a similar relation to the $L_{\rm
X}$(HMXB)-SFR-$Z$ plane, thus implicating HMXBs as a potential source of the
He~{\small II} ionization.  In an attempt to extrapolate HMXB spectral models into the
EUV, \citet[][hereafter, Sen20]{Sen2020} used multicolor accretion disk models
to show HMXBs would fall short of producing the requisite levels of ionizing
photons needed to power He~{\small II} $\lambda$4686 in low-metallicity
galaxies, unless $L_{\rm X}$/SFR~$\simgt 10^{42}$~\lum, a value that is
$\approx$100 times that observed for typical galaxies.  Also, it has been noted
that the spatial extent of He~{\small II} $\lambda$4686 and \xray\ emission do
not always coincide \citep[e.g.,][]{Keh2021}, arguing against a direct causal
connection.  However, in more recent works it has been argued that the
extrapolations into the EUV from \citetalias{Sen2020} are likely to be
unrealistic for fits to actual ULXs \citep[e.g., ][hereafter, Sim21; see
below]{Sim2021}, and the effects of ULX beaming may not yield direct spatial
coincidence between ULXs and the nebulae that they irradiate \citep{Ric2021}.

In Figures~\ref{fig:ext}$b$ and \ref{fig:ext}$c$, we show our intrinsic SED
model, after the removal of obscuration and nebular/dust emission from our
observed models.  From Figure~\ref{fig:ext}$b$, our extrapolation suggests that
the EUV flux is dominated by \xray\ emission from hot gas, with significant
contributions from stellar emission near the He~{\small II} ionization
potential at 0.054~keV (denoted as a vertical line).  For our adopted model,
the HMXB (or ULX) contribution is a factor of $\approx$2--10 times below that of the hot
gas across the EUV range.  However, it is important to note that the HMXB versus hot
gas normalization ratio is highly uncertain and subject to large galaxy-to-galaxy statistical fluctuations, especially for galaxies with low SFRs, due to stochastic sampling of the HMXB XLF and variations in ULX spectra.  Stochastic fluctuations are expected to
be amplified with decreasing metallicity as the bright end of the HMXB XLF flattens
(see, e.g., \S5.1 of \citetalias{Leh2021} for details).  As
such, the relative ULX-to-hot-gas components could easily fluctuate by an order
of magnitude or more for galaxies with SFR~$\simlt 0.1$~\sfr, and HMXBs (ULXs,
in particular) may indeed play an important role in the overall EUV photon
budget.  

To provide context to our constraint, in
Figure~\ref{fig:ext}$c$ we have overlaid the SED of \citetalias{Sen2020} ({\it
magenta dash-dot curve\/}) normalized to a value of $L_{\rm
X}$(HMXB)/SFR~=~$10^{40}$~\lum~(\sfr)$^{-1}$, comparable to the average value
of our sample.  For a fixed $L_{\rm X}$(HMXB)/SFR, extrapolation of
the \citetalias{Sen2020} curve results in a factor of $\approx$10--100 times
lower intensity than our model produce from 0.054--0.5~keV where He~{\small II}
ionizing photons are important.  
If we boost the $L_{\rm X}$(HMXB)/SFR value of our model by a factor of 10,
while holding the hot gas component fixed (see dashed curve in
Figure~\ref{fig:ext}$c$), the ionizing photon rate would consistently exceed
the \citetalias{Sen2020} rate by a factor of $\approx$100 across the EUV.  Such
a large positive fluctuation in $L_{\rm X}$(HMXB)/SFR to
just above $10^{41}$~\lum~(\sfr)$^{-1}$ is readily observed in dwarf galaxies, and is
comparable to that observed for I~Zw~18, which has been studied extensively for
its He~{\small II} emission signatures \citep[e.g.,][]{Keh2021,Ric2021}.  

For further comparison, we have displayed in Figure~\ref{fig:ext}$c$ the SED
models considered by \citetalias{Sim2021}, who evaluated the impact of various
ULX model extrapolations on high-ionization nebular emission lines He~{\small
II} $\lambda$4686 and [Ne~{\small V}] $\lambda$3426.  Three of the
\citetalias{Sim2021} models include a stellar component of age 1~Myr, based on
BPASS models (``STE'') that is combined with three ULX models, including: (1) an
accretion disk irradiated by an inner corona
\citep[``DIS'';][]{Gie2009,Ber2012}; (2) thermal photons produced in the inner
region of an accretion disk that are Comptonized by material undergoing
relativistic bulk motion \citep[``BMC'';][]{Tit1997,Ber2012}; and (3) a
Comptonization model \citep[][]{Pou1996} with a multicolor accretion disk that
was used to fit NGC 5408 X-1, a ULX surrounded by a He~{\small II} nebula
\citep[``KAA'';][]{Kaa2009}. 
In addition to these three ULX-based models, \citetalias{Sim2021} analyze the
optical--to--X-ray SED model presented in \citet[``LEB'';]{Leb2017}, which
includes the combination of stellar plus accretion disk components that
successfully model the observed optical and \xray\ spectra of I~Zw~18 and provide
the requisite ionizing flux to power the observed He~{\small II} emission in
that galaxy.

For the three ULX models, \citetalias{Sim2021} varied the ratio of {\it
intrinsic} $L_{\rm X}^{\rm int}$/SFR (where SFR is estimated using the
1500~\AA\ luminosity) and used photoionization modeling to show that He~{\small
II} $\lambda$4686 and [Ne~{\small V}] $\lambda$3426 could be produced at the
observed levels for values $L_{\rm X}^{\rm int}$/SFR~$\simgt 10^{40}$~\lxsfru\
for the DIS model and $L_{\rm X}^{\rm int}$/SFR~$\simgt$~$3 \times
10^{40}$~\lxsfru\ for the BMC and KAA models.  For illustrative purposes, the
\citetalias{Sim2021} models displayed in Figure~\ref{fig:ext}$c$ have been
normalized to $L_{\rm X}^{\rm int}$/SFR~$= 3 \times 10^{40}$~\lxsfru\, a level
that is sufficient to power the He~{\small II}.  These models produce
comparable levels of ionizing photons in the EUV and similar values of {\it observed} $L_{\rm X}$/SFR
(i.e., $\beta_{\rm HMXB}$) to our nominal model.  Furthermore, we find that
observed statistical fluctuations of $\beta_{\rm HMXB}$ that are a factor of
$\approx$10 above the average value can produce enhanced levels of ionizing emission over those of the
\citetalias{Sim2021} BCM and KAA models with $L_{\rm X}^{\rm int}$/SFR~$= 3 \times
10^{40}$~\lxsfru.  Given the large uncertainties in extrapolating to the
EUV, it is therefore not possible to exclude \xray\ emitting sources as potentially
important sources of producing these high-ionization emission lines.

%
\section{Summary}\label{sec:sum}
%

We have combined UV--to--IR data and new \chandra\ observations of a sample of
relatively low metallicity ($Z \approx 0.3 Z_\odot$) star-forming galaxies,
located at $D \approx$~200--450~Mpc, to characterize the population average
0.5--8~keV spectral shape and normalization per SFR.  We spectrally
disentangled the relative HMXB and hot gas emission components of the spectrum
and evaluated how $L_{\rm X}$/SFR for both components in these low-metallicity
galaxies compares with more typical (and higher metallicity) populations in the
nearby Universe.  Our findings are summarized as follows:

\begin{enumerate}

\item Our $Z \approx 0.3 Z_\odot$ galaxy sample was constructed to have
relatively high SFR values (0.5--15~\sfr) and low stellar masses
($\log M_\star/M_\odot =$~8.0--9.3) with very high sSFRs (3--9~Gyr$^{-1}$) to ensure the \xray\ emitting populations are
dominated by HMXBs and hot gas, which are associated with young star-formation.  As such, our sample is biased towards young stellar ages, and the HMXB populations may be associated with populations that are younger and more luminous than those presented in the literature with similar galaxy metallicities.

\item \chandra\ observations were conducted to obtain sufficient \xray\ counts
for constraining the sample-average 0.5--8~keV spectral shape. The relatively
large size of our sample, compared to past studies of low-metallicity galaxies,
minimizes the expected effects of stochastic variations on the average scaling relation.

\item Using the full \chandra\ data set, we constructed a global SFR-scaled
spectral model that consisted of an absorbed two-temperature plasma (hot gas)
with $kT \approx 0.2$~and 0.7~keV plus an absorbed power-law component (HMXBs)
with photon index $\Gamma \approx 1.8$.  Our global model provides a
statistically acceptable description of the \chandra\ spectral data set, taken
as a whole, with the majority of the galaxies being well fit on an individual
basis.  We identify only a single potential AGN (J101815.1+462623.9), which
contains an excess of counts over our model near 6--7~keV, plausibly indicating
the presence of a heavily-obscured/Compton-thick AGN.  

\item Our global model provides a means for extracting scaling relations
between SFR and the \xray\ emission from hot gas and HMXBs separately.  We find
that the SFR-scaled HMXB \xray\ luminosity (i.e., $L_{\rm 0.5-8~keV}^{\rm
HMXB}$/SFR) is elevated for our $Z \approx 0.3 Z_\odot$ galaxies compared with
higher-metallicity galaxy samples.  Specifically, we find $L_{\rm
0.5-8~keV}^{\rm HMXB}$/SFR~$= 40.19 \pm 0.06$, which is $\approx$4 times higher
than the solar-metallicity scaling relation.  This value is also somewhat elevated (by a factor of $\approx$1.2--1.5) compared with extrapolated $L_{\rm X}$-SFR-$Z$ relations in the literature, which we speculate may be due to our sample having relatively low intrinsic HMXB obscuration and elevated HMXB emission from relatively young and \xray-luminous HMXB populations.

\item For the hot gas component, we find a SFR scaling of $L_{\rm
0.5-2~keV}^{\rm gas}$/SFR~=~$39.58_{-0.28}^{+0.17}$ for our low-metallicity
sample.  We use our constraints, along with measurements from local galaxies
with comparable SFR values \citep{Min2012b}, to show that $L_{\rm
0.5-2~keV}^{\rm gas}$/SFR declines with metallicity.  The level of this trend
is consistent with increased soft \xray\ absorption from metal lines and edges
with increasing metal abundances for a fixed hydrogen column density.  However,
this trend is only weakly constrained, and more detailed studies are required
to both verify the trend and uncover its physical nature.

\item Using our \xray\ global model, along with stellar, nebular, and dust
emission and absorption models of the UV--to--IR data, we construct both
emergent (observed) and intrinsic IR--to--\xray\ SFR-scaled SED models.  We
show that by making reasonable interpolations into the EUV, our constraints on
low-metallicity galaxy emergent emission are consistent with recent HERA
constraints on \xray\ emission from high-redshift galaxies, which are expected
to contain low-metallicity galaxies.  We also find that our intrinsic SED model
produces significant EUV emission, and with modest fluctuations in the ULX
population luminosity, which are commonly observed in dwarf galaxies, the EUV
emission can exceed that required to power observed He~{\small II} nebulae.

\end{enumerate}

\vspace{0.1in} {\large {\it Acknowledgements:}} We thank the anonymous referee for their helpful comments on the manuscript.  We thank Charlotte Simmonds for
kindly providing spectral energy distribution models to compare with our
results.  We gratefully acknowledge financial support from \chandra\ X-ray
Center (CXC) grant GO0-2J076A (B.D.L., R.T.E., A.B.) and NASA Astrophysics Data
Analysis Program 80NSSC20K0444 (B.D.L.,R.T.E.).  Some of the material is based
upon work supported by NASA under award number 80GSFC21M0002 (A.B.).  K.G. acknowledges support by the NASA Postdoctoral Program at Goddard Space Flight Center, administered by Oak Ridge Associated Universities under contract with NASA.  A.M.
gratefully acknowledges support by the European Research Council (ERC) under
the European Union's Horizon 2020 research and innovation programme (grant
agreement No 638809 -- AIDA). The results presented here reflect the authors'
views; the ERC is not responsible for their use.

This research has made use of the NASA/IPAC Extragalactic Database (NED),
which is operated by the Jet Propulsion Laboratory, California Institute of Technology,
under contract with the National Aeronautics and Space Administration.

\vspace{5mm}

\facilities{{\it Chandra}, {\it GALEX}, PanSTARRS, Sloan, 2MASS, {\it WISE}}

\software{CIAO \citep[v4.13][]{Fru2006}, {\ttfamily DS9} \citep[][]{Joy2003}, {\ttfamily Lightning} \citep{Euf2017,Doo2021}, {\ttfamily P\'EGASE} \citep{Fio1997}, {\ttfamily Sherpa} \citep[v4.13.0][]{Bur2021}, {\ttfamily XSPEC} \citep{Arn1996}}

\bibliography{mybib.bib}{}
\bibliographystyle{aasjournal}

\end{document}